\documentclass[aps,preprint,onecolumn,showpacs]{revtex4}

\usepackage{amsmath}
\usepackage{amssymb}
\usepackage{dcolumn}
\usepackage{bm}
\usepackage{graphicx}

\def\Tc {$T_{\text{c}}$}

\def\dT {$\Delta T$}
\def\dsdh {$(\partial{s}/\partial{H})_{T}$}
\def\mt {$T_{\text{s}}(\partial{s}/\partial{H})_{T}$}

\def\Hl {$H_{1}$}
\def\Hon {$H_{\text{on}}$}
\def\Hh {$H_{\text{hyst}}$}
\def\Hp {$H_{\text{p}}$}
\def\Ho {$H_{\text{0}}$}

\def\Hab {$H_{\text{c2}}$}
\def\Hav {$H^{\text{av}}_{\text{c2}}$}
\def\Hup {$H^{\text{up}}_{\text{c2}}$}

\begin{document}

\title{Magnetocaloric Studies of the Peak Effect in Nb}

\author{N. D. Daniilidis}
	\email[Corresponding author: ] {nikos@brown.edu}
\author{I. K. Dimitrov}
\author{V. F. Mitrovi{\'c}}
\author{C. Elbaum}
\author{X. S. Ling}
	\email[Corresponding author: ] {xsling@brown.edu}
\affiliation{Department of Physics, Brown University, Providence, Rhode Island 02912}%

\date{\today}

\begin{abstract}
We report a magnetocaloric study of the peak effect and Bragg
glass transition in a Nb single crystal. The thermomagnetic 
effects due to vortex flow into and out of the sample 
are measured. The magnetocaloric signature of the peak effect 
anomaly is identified. It is found that the peak effect disappears in 
magnetocaloric measurements at fields significantly higher than those 
reported in previous ac-susceptometry measurements. Investigation of 
the superconducting to normal transition reveals that the disappearance 
of the bulk peak effect is related to inhomogeneity broadening of the 
superconducting transition. The emerging picture also explains the 
concurrent disappearance of the peak effect and surface superconductivity, 
which was reported previously in the sample under investigation. Based on 
our findings we discuss the possibilities of multicriticality associated 
with the disappearance of the peak effect.
\end{abstract}


\maketitle


\section{\label{Sec1}Introduction}

One very fascinating result in condensed matter physics
in recent decades is the understanding that, in spite of early
predictions,\cite{IM} the long-range topological order associated
with broken continuous symmetries can survive in systems with
random pinning.\cite{natt, GL} In bulk type-II superconductors
with weak point-like disorder the existence of a novel Bragg
glass phase has been predicted.\cite{GL} This reaffirmed 
experimental facts known since the 1970s, that vortex lattices 
in weak-pinning, bulk, type-II superconductors can produce 
pronounced Bragg peaks in neutron diffraction.\cite{Christen} 
Recent experiments\cite{Ling, Troy} have shown that a genuine
order-disorder transition occurs in vortex matter. This transition 
appears to be the underlying mechanism of the well-known anomaly of the
peak effect\cite{PE} in the critical current near
\Hab\hspace{0.01cm}.  However there are still many outstanding
issues concerning the Bragg glass phase boundary and the nature of
the disordered vortex state above the peak effect.

\begin{figure} [ht]
\centering
\includegraphics[scale=0.5] {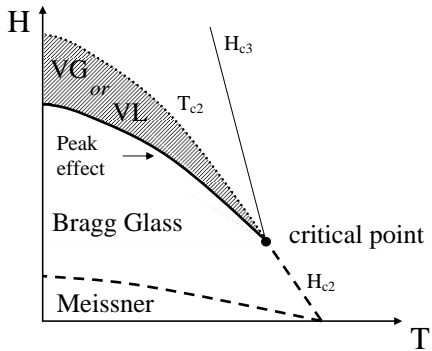}
\caption{\footnotesize{A sketch of the phase diagram of the Nb
single crystal from Park \textit{et al.}\cite{Park}. Note the 
definitions of distinct $T_{\text{c2}}$, $H_{\text{c2}}$ and 
peak effect lines. }}
\label{Fig1}
\end{figure}

Previous studies in a Niobium single crystal have revealed an 
intriguing picture of the peak effect in weakly-pinned conventional 
superconductors. Neutron scattering has shown that a vortex lattice 
order-disorder transition occurs in the peak effect region. This 
transition shows hysteresis and is believed to be first order, 
separating a low temperature ordered phase from a high temperature 
disordered one.\cite{Ling} The hysteresis was not observed 
across the lower field part of the superconducting-to-normal phase
boundary. Magnetic ac-susceptometry showed that 
at lower fields the peak effect disappears as well, indicating further 
connection between the peak effect and the order-disorder 
transition.\cite{Park} In addition, the line of surface 
superconductivity, $H_{\text{c3}}$, terminates in the vicinity
of the region where the peak effect disappears. This picture is 
summarized in Fig.\,\ref{Fig1}.

It was thus proposed that the peak effect is the manifestation of 
a first-order transition which terminates at a multicritical point 
(MCP) where the peak effect line meets a continuous, Abrikosov 
transition, \Hab. The MCP would be bicritical if a third line of 
continuous transitions ends there. The transition lines considered 
as a possible third candidate were a continuous vortex glass 
transition, $T_{\text{c2}}$, and the line of surface superconductivity. Alternatively, the MCP 
would be tricritical if the disordered phase is a pinned liquid 
with no high field transition into the normal state.\cite{Park} 
Subsequently, the disappearance of the peak effect at low fields 
has  also been reported in other systems.\cite{Adesso,Jaiswal}

Thermodynamic considerations\cite{Park} suggest that the MCP 
is likely a bicritical point.  Since bicriticality implies
the existence of competing types of order in the vortex system the
question of which of the two lines, $T_{\text{c2}}$ or 
$H_{\text{c3}}$, is relevant to the bicritical point has major
importance. Its answer will provide insight into the disordered 
vortex state above the peak effect and the disordering transition 
itself. Evidently, the possible relevance of surface 
superconductivity to the destruction of bulk Bragg glass ordering,
and hence the existence of the multicritical point, cannot be 
dismissed \textit{a priori}. In fact it is well known that surface 
premelting plays an important role in solid-liquid transitions.\cite{SurfMelt} 

This issue could be resolved by repeating the ac magnetic susceptibility 
measurements after having suppressed surface superconductivity with 
appropriate surface treatment. Efforts to nondestructively achieve 
this, e.g. by elecroplating the sample surface with a ferromagnetic 
layer, proved unfruitful, possibly due to high oxygen content of the 
surface. To address the problem, we performed 
measurements of magnetocaloric effects on the Nb single crystal 
studied by Ling \textit{et al.}\cite{Ling} and Park 
\textit{et al.}\cite{Park} 

Here we report a study of the peak effect using a magnetocaloric technique. Clear features
associated with the peak effect, have been identified in the magnetocaloric data. We find
that the superconducting to normal transition shows inhomogeneity broadening 
at all fields. The peak effect in the bulk critical current is found to 
disappear when it enters into the inhomogeneity broadened 
transition region. It is concluded that the concurrent disappearance of surface 
superconductivity and the peak effect is likely due to interference of surface 
superconductivity with the ac-susceptibility measurements. Nevertheless, the 
location of the MCP, as determined from magnetocaloric measurements, is in close 
vicinity to the location previously reported.\cite{Park} The transition across 
$T_{\text{c2}}$ seems to be the same as that at $H_{\text{c2}}$, and surface 
superconductivity plays only a coincidental role in the critical point. This 
result suggests that the disordered vortex state, which is represented by the 
shaded part in Fig.\,\ref{Fig1}, is a distinct thermodynamic phase. Bicriticality 
implies that this phase has an order parameter competing with that of the Bragg 
glass phase. We also discuss an alternative scenario, in which no critical point 
exists at any finite field or temperature.

The paper is organized as follows: In Sec.\,\ref{Sec2} we review the 
basic principle of magnetocaloric measurements, and give the experimental 
details of the sample and the technique used. In Sec.\,\ref{Sec3} we
present our main results, discuss the consequences of irreversible and 
nonequilibrium effects on magnetocaloric measurements, and proceed to 
interpretation of the data and estimates of sample properties. Finally 
we summarize our findings and conclusions, and propose experimental work 
necessary to address the issues raised.

\section{\label{Sec2}Experimental}

\subsection{\label{Sec2a}Basic principle of magnetocaloric measurements}

\subsubsection{\label{Sec2a1} Heat flow in magnetocaloric measurements}

In studies of vortex phases in bulk superconductors, various experimental 
techniques provide complementary pieces of information. Combining these in a 
consistent picture is a non-trivial task. Magnetic ac-susceptibility 
measurements are sensitive to screening currents, and thus to the location of 
peak effect features, but are not suitable for the identification and 
study of the mean field \Hab\hspace{0.01cm} transition itself.\cite{Park} 
Commercial magnetometers are not suited for study of large samples. 
Calorimetric\cite{Lortz, Daniilidis} and ultrasonic attenuation\cite{Shapira, Dimitrov} 
measurements determine the upper critical field where bulk condensation 
of Cooper pairs occurs, but it remains unclear under 
what circumstances they also provide a peak effect signature. Moreover the 
combination of information obtained with different techniques has to rely on 
thermometer calibration issues. Furthermore, the dynamical measurements suffer 
from thermal gradients in the studied samples. Magnetocaloric measurements overcome 
these difficulties because they are sensitive to both the presence of bulk 
superconductivity and to dynamical, flux-flow related effects. Moreover they can 
be performed in quasi-adiabatic conditions, where virtually no thermal gradients 
are present in the sample. Finally, they can be easily performed using a common
calorimeter.
 
The magnetocaloric effect is a special case of thermomagnetic effects in the 
mixed-state of superconductors which have long been known and investigated.\cite{Thermomagnetic} 
These arise from the coexistence of the superconducting condensate which is not involved 
in entropy-exchange processes for the superconductor, and the quasiparticles, 
localized in the vortex cores, which are entropy carriers. Due to the presence 
of localized quasiparticles in the vortex cores,  vortex motion results 
in entropy transport, which causes measurable thermal effects. 
Specfically, during field increase, new vortices are created at the edge of the sample, 
quasiparticles inside the vortices absorb entropy from the atomic lattice and 
cause quasi-adiabatic cooling of the sample. Conversely, during field 
decrease, the exiting vortices release their entropy to the atomic lattice, causing 
quasi-adiabatic heating. 

Typically, in an experiment, the amount of entropy carried by the 
vortices entering or leaving the sample  is not the exact thermodynamic equilibrium 
vortex entropy. The reason is that the vortex assembly is not equilibrated 
due to pinning as well as metastability associated with the underlying first-order phase 
transition at the peak effect. In practice, the magnetocaloric cooling and heating 
due to vortex entry and exit can be understood formally in terms of the actual entropy 
exchange between the vortex and atomic lattices.  A superconductor subject to a changing 
magnetic field and allowed to exchange heat with the environment, undergoes a temperature 
change.\cite{adiabatic} This process is described by the relation: 
\begin{equation} \label{Eq1}
			dQ_{abs}/dt=n\,T_{\text{s}}\,(\partial{s}/\partial{H})_{T}\,dH/dt+
			n\,C_{\text{s}}\,dT_{\text{s}}/dt
\end{equation}
\noindent where $dQ_{abs}/dt$ is the net rate of heat absorption, positive 
for absorption of heat by the sample, $n$ is the molar number of the superconductor, 
$T_{\text{s}}$ its temperature, $T_{\text{s}}\,(\partial{s}/\partial{H})_{T}$ 
the molar magnetocaloric coefficient, and $C_{\text{s}}$ the specific heat of 
the superconductor. The magnetocaloric term in the equation induces 
temperature changes. These are smeared out by the last term, describing the 
effect of specific heat. Nevertheless, in practice this last term is constrained 
to be negligible when magnetocaloric effects are measured. In our measurements, 
we use very low field ramp rates ($dH/dt$), which result in very low temperature 
change rates ($dT_{\text{s}}/dt$) causing the last term to be negligible. 

A schematic of our setup is represented in Fig.\,\ref{Fig2}a.
In this setup, absorption or release of heat from the sample results in 
minute, but measurable, variation of its temperature. In a typical measurement, 
the sample temperature is first fixed at a selected value, $T_{\text{s0}}$. 
Subsequently, the field is ramped at a steady rate, resulting in quasi-adiabatic 
absorption or release of heat from the sample. The resulting sample temperature 
change is recorded. If we neglect irreversible and non-equilibrium effects, 
the temperature change of the sample (\dT) around its static value, 
$T_{\text{s0}}$, allows us to determine the molar magnetocaloric coefficient 
\mt\hspace{0.01cm} by use of:
\begin{equation} \label{Eq2}
			-G_{\text{link}}\,\Delta T=n\,T_{\text{s}}\,(\partial{s}/\partial{H})_{T}\,dH/dt
\end{equation}
\noindent where the sample temperature 
$T_{\text{{s}}}=T_{\text{s0}}+\Delta T$, differential thermal conductance of 
the heat link $G_{\text{link}}$, molar number for the sample $n$, and field 
ramp rate $dH/dt$ are all independently measured. The latter is positive for 
increasing fields, negative for decreasing fields.

\subsubsection{\label{Sec2a2}Irreversible and non-equilibrium effects}

In deriving the above equations we assumed that heat is only exchanged 
between the superconductor and the environment and that the superconductor 
reaches its quasi-static state as the measurement is performed. In an actual 
experiment, non-equilibrium and irreversible effects need to be considered. 
The heat generated by the dissipative processes between the vortex lattice 
and atomic lattice leads to a modification of the left hand side of equations
\ref{Eq1}\,\&\ref{Eq2}. Non-equilibrium effects lead to a modification of the 
right hand side. While the effects of the Bean-Livingston surface 
barrier,\cite{BeanLiv} flux flow heating,\cite{FF} and the Bean critical 
state\cite{Bean} can be complex and subtle, a detailed analysis can be performed 
and it will be discussed in Sec.\,\ref{Sec3b1}. We show that the magnetocaloric 
measurements allow us to shed light into the problem of the peak effect in Nb
that would not have been possible with any other single technique. In concluding 
this section, we note that since Eq.\,\ref{Eq2} is approximate, in what follows 
we refrain from using \dsdh\, to symbolize the quantity related to the magnetocaloric 
temperature change, $\Delta T$. Instead, we choose $(ds/dH)$ to symbolize the measured
``entropy'' derivative.

\subsection{\label{Sec2b}Sample and setup}

\begin{figure} [ht]
\centering
\includegraphics[scale=0.4] {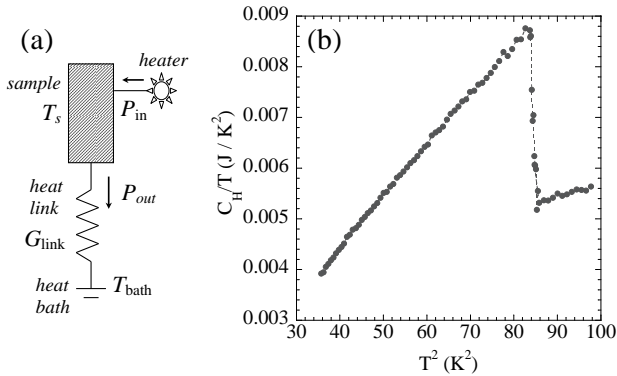}
\caption{\footnotesize{(a) Idealized heat-flow diagram of the calorimeter used. 
(b) Specific heat of Nb crystal studied in zero applied field. The small 
contribution from the addenda of the calorimeter is included. }}
\label{Fig2}
\end{figure}

We used a Nb single crystal sample with mass of $24.78\, \text{g}$ 
which was previously studied using SANS and ac-susceptometry.\cite{Ling,Park} 
It has an 
imperfect cylindrical shape (radius $0.5\, \text{cm}$, length 
$2.47\, \text{cm}$) with the cylinder axis oriented parallel to 
the [111] crystallographic direction. It has a \Tc\hspace{0.01cm} 
of $9.16\, \text{K}$ and upper critical field \Hab$(0)\approx5600\, \text{Oe}$,
as previously reported.\cite{Ling,Park} We performed a zero field 
specific heat measurement, shown in Fig.\,\ref{Fig2}b. This was 
done with the heat-pulse (relaxation) technique. From this 
measurement we obtain the  Ginzburg-Landau parameter $\kappa(0)=3.8$ which 
is higher than the previous estimate\cite{Park}. In addition we find a 
superconducting transition width of $83\,\text{mK}$. The residual 
resistivity ratio from $300\, \text{K}$ to 
$10\, \text{K}$ of this sample is measured to be 12, suggesting 
a significant amount of defects in this Nb crystal. This is 
consistent with the large $\kappa$ and upper critical field values.

Our experimental setup is a homemade calorimeter and its idealized 
heat-flow diagram is shown in Fig.\,\ref{Fig2}a. It consists of an
oxygen-free, high-purity copper can which serves the role of a
heat bath surrounding the sample. A piece of high-purity copper
wire is used as a heat link between the sample and heat bath and
the mechanical support for the sample is provided by nylon rods.
During measurements, the heat bath is maintained at a temperature
of $4.20\, \text{K}$ and a carbon glass thermometer (Lakeshore) 
monitors its temperature. A second thermometer (Cernox, Lakeshore) 
is directly attached with silver epoxy to the lower end of the 
sample for reading the sample temperature. To minimize electronic 
noise in reading the thermometer resistance, high-frequency 
sinusoidal excitation together with pulse-sensitive detection is 
used. A Manganin wire, which is non-inductively wound on
the sample and secured with Stycast epoxy, serves as a heater.  
A 50-turn high-purity copper coil is directly wound on the sample
for simultaneous ac magnetic susceptibility measurement. During 
measurements the vacuum in the calorimeter is maintained to lower 
than $1\, \text{micron}\, \text{Hg}$ by use of activated charcoal 
in thermal contact with the helium bath.

Calorimetric measurements are performed with the standard 
heat-pulse technique. The heat input to the sample is increased 
incrementally, with step duration of 60-100 sec. Every incremental 
increase of the heat input results in exponential temporal
relaxation of the sample temperature to its new equilibrium value. 
The sample heat capacity is determined through the decay time of 
the exponential. This technique offers moderate resolution and is 
not suited for studying the thermodynamics of the peak effect. 
Nevertheless, it allows us to characterize our sample. In 
Fig.\,\ref{Fig2}b, we show the data acquired in zero applied field, 
from which we measured the sample properties quoted above.
  
\begin{figure}[!t]
\includegraphics[scale=0.45] {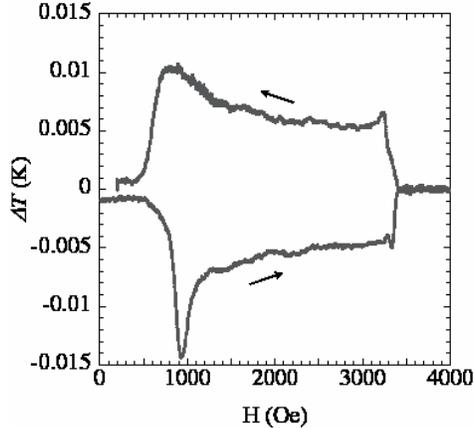}
\caption{\footnotesize{Magnetocaloric temperature variation around $T_{s0}=5.37$\,K 
$vs.$ applied field, for increasing and decreasing field. The field ramp direction is 
indicated by arrows. $dH/dt=0.92\,$Oe$\,/\,$sec was used.}}
\label{Fig3}
\end{figure}

During magnetocaloric measurements, a constant heat input, $P_{in}$, 
is supplied to 
the sample through the manganin heater, fixing the temperature at a 
selected static value, $T_{\text{s0}}$. After the sample 
temperature has stabilized to $T_{\text{s0}}$, the magnetic field
is ramped up, then down, at a constant rate. During a field ramp, 
the sample temperature as a function of field, $T_{\text{s}}(H)$, 
is recorded. The magnetocaloric signal, 
$\Delta T(H)$, has to be measured with respect to a 
\textit{field dependent}, static sample temperature \textit{reading}: 
$T_{\text{s0}}=T_{\text{s0}}(H)$. This is the 
thermometer reading obtained at a given field, $H$, in the absence 
of field ramping. In other words one has to determine
$\Delta T(H)=T_{\text{s}}(H)-T_{\text{s0}}(H)$. An example of raw 
data, $\Delta T$ $vs.\;H$, is shown in Fig.\ref{Fig3}.

The field dependence of the $T_{\text{s0}}(H)$ thermometer reading 
is a result of the following two effects: First, the magnetoresistance 
of the thermometer used. Second, a changing temperature gradient across 
the sample, as its field dependent thermal conductivity changes. This 
gradient can become considerable in the Meissner state for the highest 
measured temperatures 
($\approx 8 \,\text{mK}\,/\,\text{cm}$ at $8.33\, \text{K}$). Nevertheless, 
it is negligible (less than $0.5 \,\text{mK}\,/\,\text{cm}$) in the peak 
effect region which is the main focus of this paper. We stress here 
that the thermal gradient is a result of the external heat input $P_{in}$, 
not the magnetocaloric effect. 

A possible variation of the heat input with applied field, which could be 
due to magnetoresistance in the heater, was investigated. It was 
verified that the heat input from the manganin heater does not 
vary by more than 1 part in $10^{4}$ for the entire field range 
covered in any of our measurements. 

The differential thermal conductance of the heat link between the sample 
and heat bath, $G_{\text{link}}$ to be used in Eq.\,\ref{Eq2} above,  
is also measured at all temperatures of interest, for the entire range of 
applied fields. This is done through application of a small, pulsed heat input 
($P_{in}\,\pm\,\delta P_{in}$) to the sample and recording of the resulting temperature 
change, at different applied fields. The link conductance is found to vary smoothly 
by no more than 1 part in 100 for the entire field range studied. This variation is
insignificant compared to the random noise in the raw $\Delta T$ data, and shall not 
be considered further. 

Finally, the 
field ramp rate is measured through a resistor connected in series with the 
magnet coil, which allows us to monitor the current through the magnet. For all of 
the measurements presented here the ramp rate is $0.92\, \text{Oe/sec}$, which is 
the lowest possible with our system. At higher ramp rates, for example 
$1.87\, \text{Oe/sec}$, giant flux jumps occur in the sample.

\section{\label{Sec3} Results and discussion}

\subsection{\label{Sec3a}Magnetocaloric results}

\subsubsection{\label{Sec3a1}Main features in field scans}

\begin{figure} [!tb]
\includegraphics[scale=0.35] {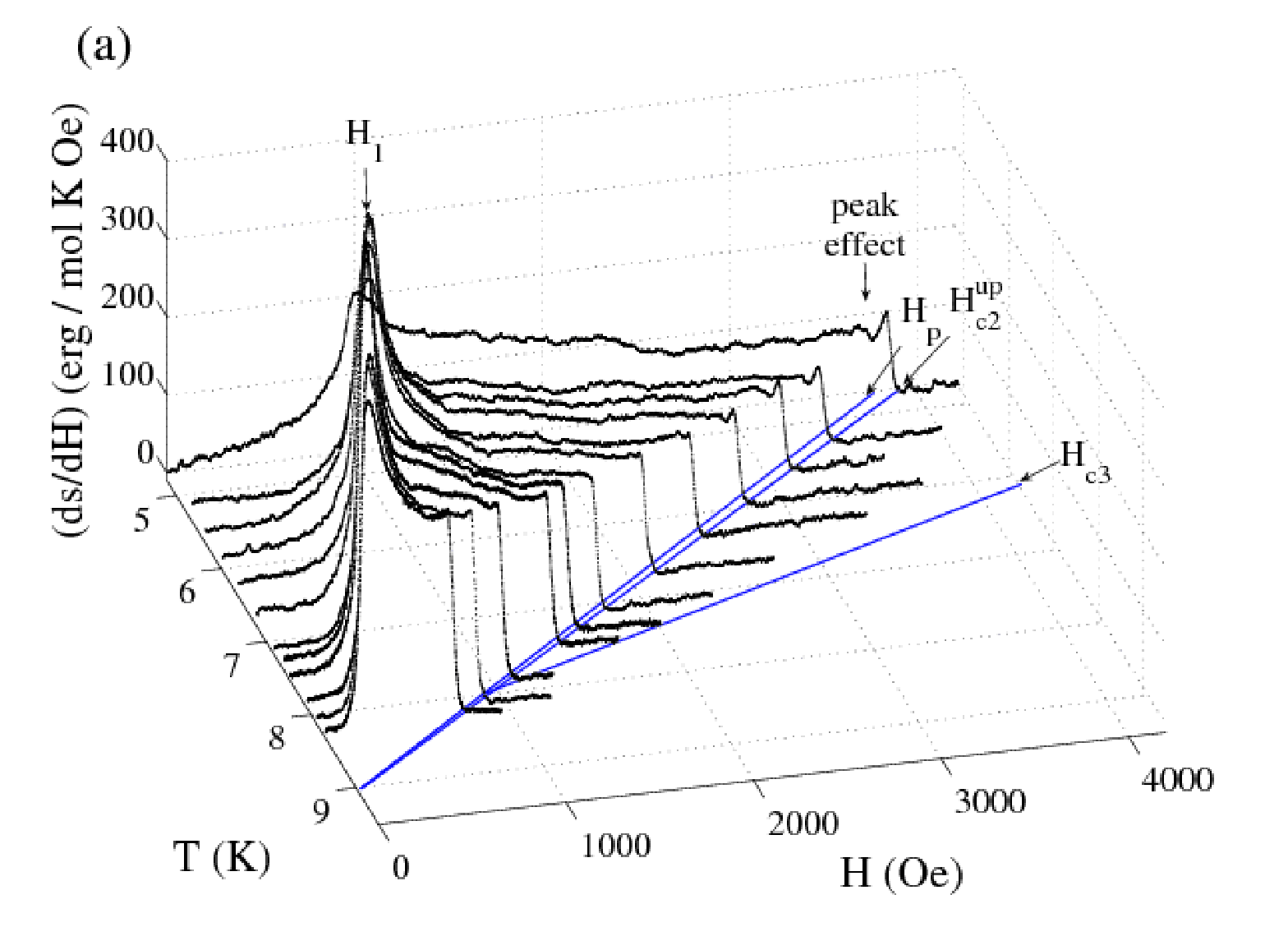}
\includegraphics[scale=0.35] {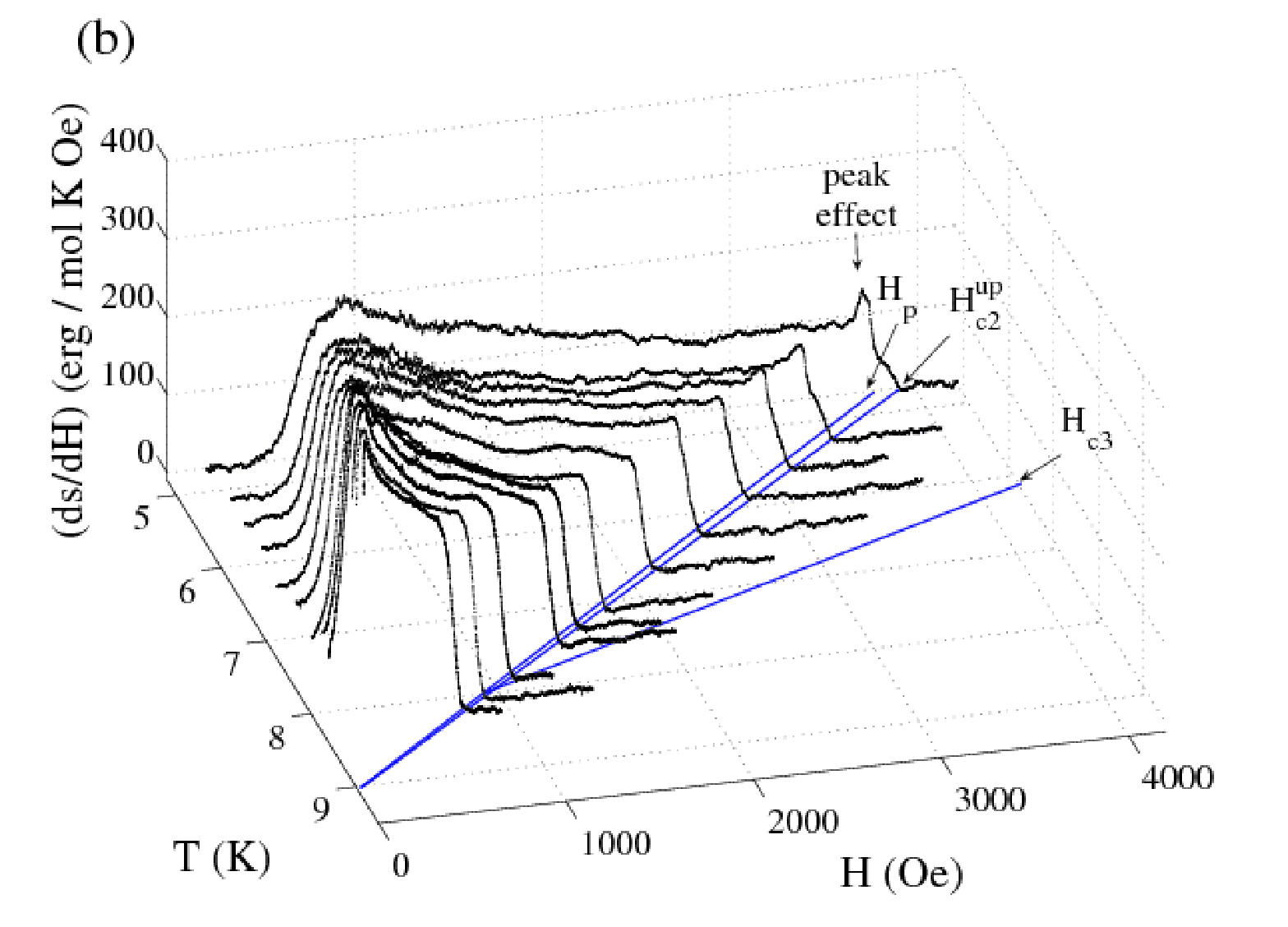}
\includegraphics[scale=0.45] {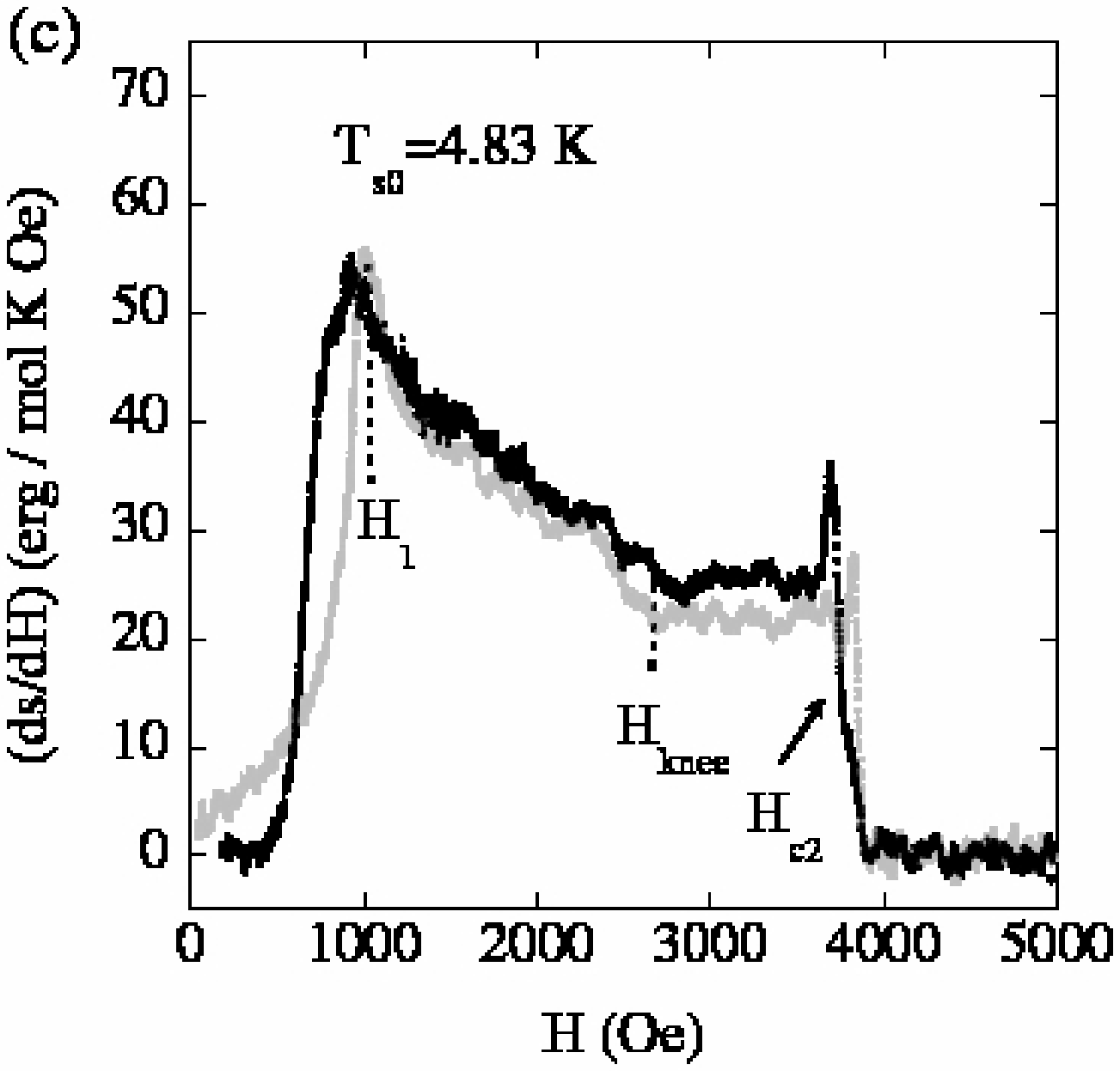}
\caption{\footnotesize{\textit{(color online)} Results of magnetocaloric 
measurements on (a) increasing and (b) decreasing field. Also shown (in 
blue) are the loci traced on the $H-T$ plane by \Hp, \Hup, and 
$H_{\text{c3}}$.  The $H_{\text{c3}}$ line is drawn according to Park 
\textit{et al.}\cite{Park}. (c) Full field scan at 
$T_{\text{s0}}=4.83\,\text{K}$. Grey: Increasing field. Black: Decreasing 
field. The fields $H_{1}$ and $H_{\text{knee}}$, and the region of the 
upper critical field, where the peak effect occurs, are marked.}} 
\label{Fig4}
\end{figure}

In Fig.\,\ref{Fig4}a\,\&\,b, we summarize the 
molar entropy derivative, $(ds/dH)$, measurements on increasing 
(a) and decreasing (b) fields, at different temperatures.  
$(ds/dH)$ is calculated from the  $T_{\text{S}}(H)$ data following 
the procedure outlined above. 

As indicated for the lowest temperature curve, $4.83\, \text{K}$ 
(the upper most curve in Fig.\,\ref{Fig4}a) several important 
features deserve attention. On increasing fields a peak occurs at 
low field. This is marked by \Hl. It corresponds to the lowest 
field for vortex entry through a surface barrier. Its locus on the 
$H-T$ plane closely follows the
thermodynamic field, but occurs slightly lower. This behavior is 
expected for a sample with finite demagnetizing factor and 
mesoscopic surface irregularities.\cite{deGennes} No corresponding peak is present 
on decreasing fields. Rather, a smoother increase of $(ds/dH)$
occurs as the field is lowered, before entry of part of the sample into the 
Meissner state where the magnetocaloric signal vanishes, as seen in
figures \ref{Fig3}\,\&\,\ref{Fig4}b. 

In intermediate fields, we identify a novel feature which was not observed in 
previous magnetic susceptibility studies.\cite{Park} This appears as a knee
in $(ds/dH)$, which shows larger negative slope as a function of field 
for fields lower than $H_{\text{knee}}$, as illustrated in Fig.\,\ref{Fig4}c. The feature 
is the same for both field-ramp directions. With our setup we can identify the 
$H_{\text{knee}}$ feature up to $7.41\,\text{K}$. As the temperature is 
increased, the region between $H_{1}$ and $H_{\text{c2}}$ narrows and it 
becomes increasingly difficult to discern $H_{\text{knee}}$. Thus it is unclear 
how this new feature terminates, i.e. whether it ends on the \Hab\hspace{.01cm} 
line at around $H=1000\,\text{Oe}$, or if it continues to lower fields. 

At high field, across $H_{\text{c2}}$, the equilibrium mean-field theory 
of Abrikosov predicts a step function for $(\partial m/\partial T)_{H}$.\cite{deGennes2} 
Thus, in the simplest picture one would expect a simple step function for 
the molar entropy derivative \dsdh\hspace{0.01cm} at $H_{\text{c2}}$.  
Instead, we observe that across the peak-effect regime, complex features 
of valley and peak appear in $(ds/dH)$ below the field marked \Hup. The valley 
in $(ds/dH)$ corresponds to the peak effect, as seen in Fig.\,\ref{Fig4}. 
Similar features appear in decreasing field. At fields above the peak 
effect, the disappearance of the magnetocaloric effect marks the upper 
critical field, \Hab. As we will soon discuss, our magnetocaloric 
measurements indicate that the upper critical field shows 
inhomogeneity broadening, in agreement with the zero field calorimetric 
data mentioned in section \ref{Sec2}\,C. In Fig.\,\ref{Fig4} we mark with 
\Hup\hspace{0.01cm} the upper end of the upper critical field. This is the
value of field at which the magnetocaloric signal in the mixed state exceeds
noise levels. Our technique is not sensitive to $H_{\text{c3}}$ effects and the data are 
featureless above \Hup.   

\subsubsection{\label{Sec3a2}Identification of the peak effect}

To verify the identification of the peak effect in our measurements, we 
performed simultaneous magnetocaloric and ac-susceptibility measurements, 
as shown in Fig.\,\ref{Fig5}a. In the quasi-adiabatic magnetocaloric setup,
the ac-amplitude used in this procedure has to be small. For large 
amplitudes inductive heating occurs in the mixed state on 
increasing \textit{dc} field. Consequently the sample temperature 
increases rapidly by several degrees as the peak effect is crossed. 
This behavior is not surprising, but reaffirms that caution has to be 
taken when dynamic perturbations are combined with quasi-adiabatic measurements.  
We used an amplitude of $0.5\,\text{Oe}$
at $107\,\text{Hz}$ as a compromise between feasibility of the 
magnetocaloric measurement and resolution in the $\chi '$ results. The
results are shown in Fig.\,\ref{Fig5}a.

In this combined measurement we find that both the onset and the peak 
of the peak effect have corresponding features. Moreover, we find no 
clear change in $\chi '$ when the upper critical field, determined from 
the magnetocaloric measurement, is crossed. A slight change occurs in 
the slope of the $\chi '(H)$ curve across \Hab, but significant amount of 
screening, caused by surface superconductivity, remains when the bulk of 
the sample is in the normal state. This is a typical example of the inadequacy 
of ac-susceptometry in determining the upper critical field. Even with 
the use of larger ac fields the change in slope of $\chi '(H)$ turns into 
a shoulder which does not reveal the exact location or characteristics of 
the bulk superconducting transition.\cite{Park}.

\subsubsection{\label{Sec3a3}Features of the peak effect}

In light of the first-order transition underlying the peak effect, it is
tempting to interpret the peak appearing in $(ds/dH)$ at the peak effect 
as a manifestation of the entropy discontinuity of the transition. Nevertheless, 
a simpler interpretation exists in the context of critical state screening. 

\begin{figure} [ht]
\centering
\includegraphics[scale=0.45] {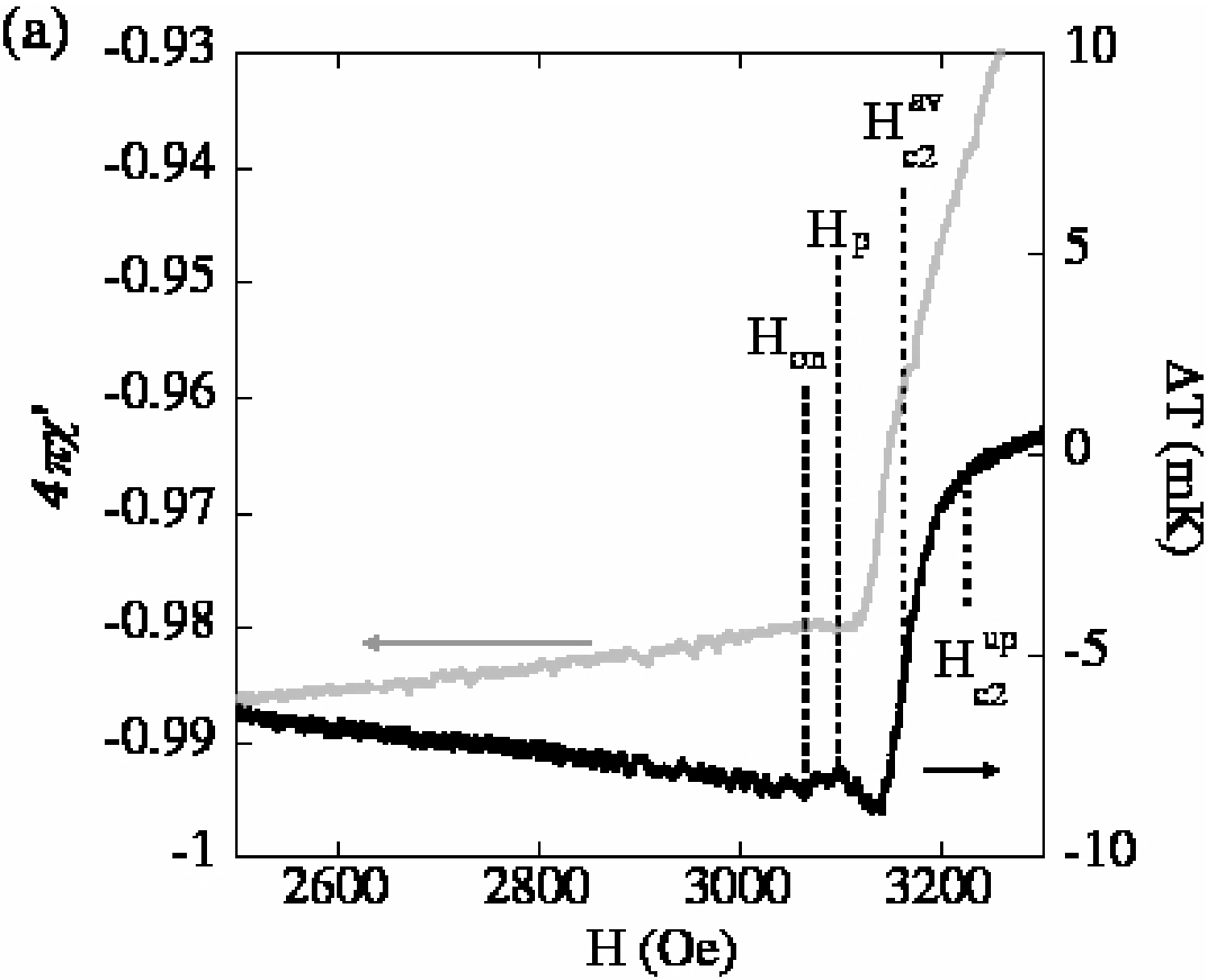}
\includegraphics[scale=0.45] {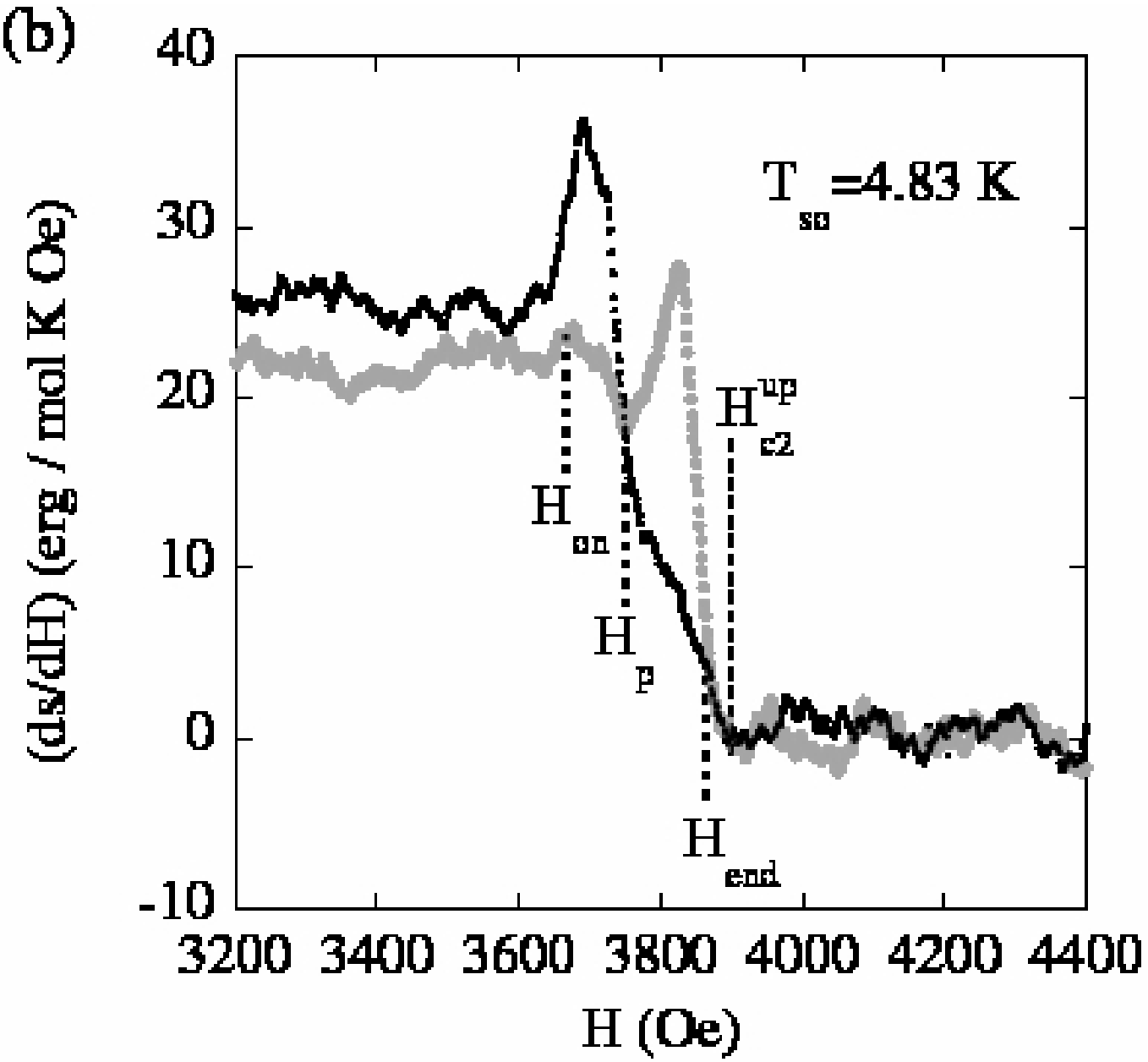}
\includegraphics[scale=0.45] {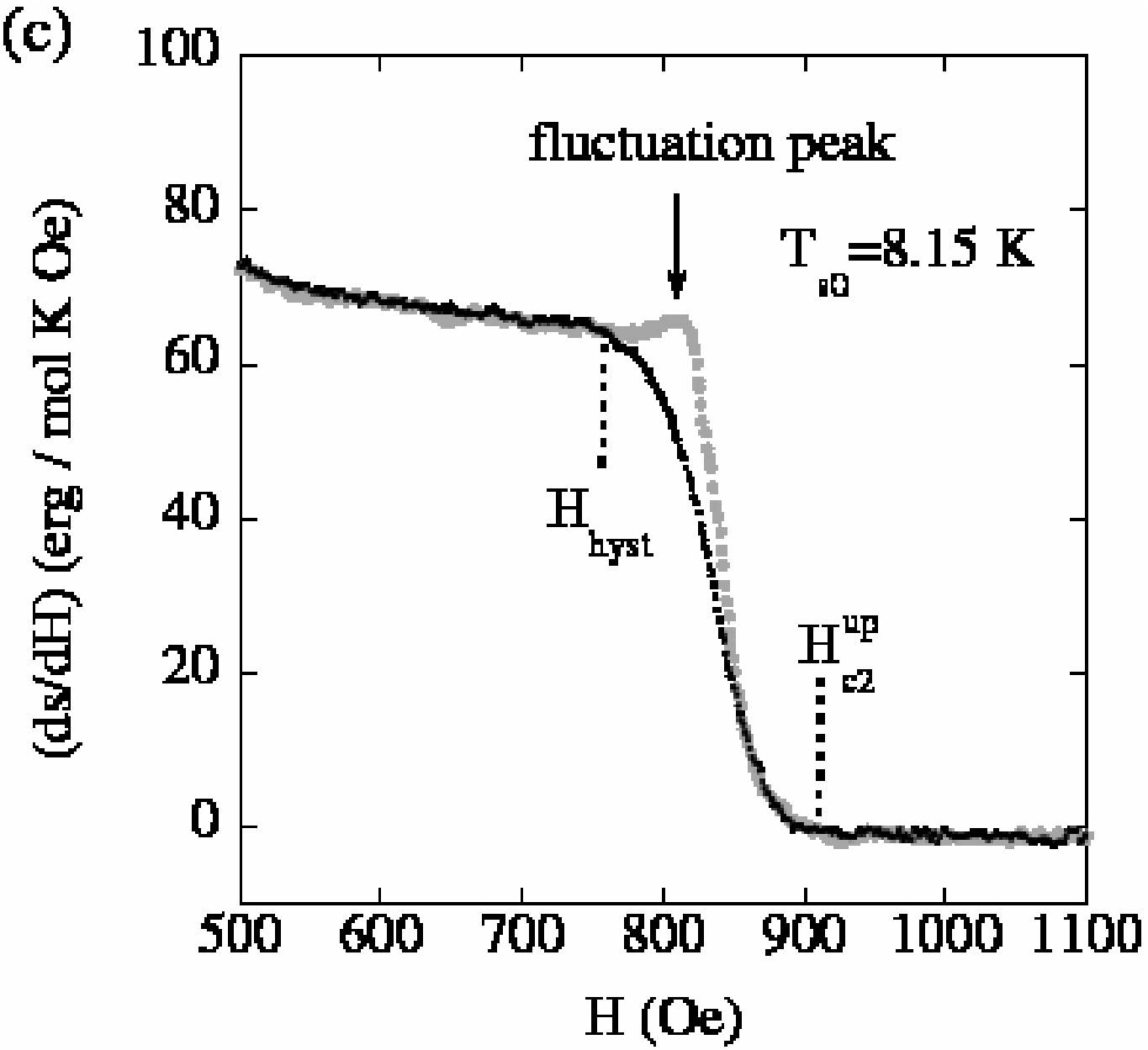}
\caption{\footnotesize{(a) Magnetocaloric temperature variations and 
ac magnetic susceptibility as a function of increasing field, 
at $T=5.76\, \text{K}$. The peak effect in $\chi'$ is not pronounced 
due to the small ac amplitude used ($h_{ac}=0.5$ Oe, $f=107$ Hz).
(b) Detail of magnetocaloric measurement in the upper critical 
field region at $T=4.83\, \text{K}$. Grey: increasing field. Black:
Decreasing field. (c) Same as b, at $T=8.15\, \text{K}$.}}
\label{Fig5}
\end{figure}

In specific, the magnetocaloric signature of the peak effect is consistent 
with critical-state\cite{Bean} induced, non-equilibrium magnetization during 
the field ramps. The magnetocaloric valley-and-peak features in the
peak-effect regime allow us to determine the positions of the onset, \Hon,
the peak, \Hp, and the end, $H_{\text{end}}$, of the peak effect.  We start with 
increasing field data. At low temperatures below $6.79\, \text{K}$ 
where the peak effect is observed, the magnetocaloric signal starts 
dropping at the onset, \Hon, of the peak effect. A minimum occurs 
in the vicinity of the peak of the peak effect, \Hp, and it is followed 
by a peak, see Fig.\,\ref{Fig5}b. This indicates slowing down, then 
acceleration of vortex entry into the sample, as the critical state 
profile becomes steep, then levels, in the peak effect region. Finally the
magnetocaloric signal gradually drops to zero in the region of the upper 
critical field.

On decreasing field a magnetically reversible region exists for 
fields between $H_{\text{end}}$ (the ``end'' of the peak effect)
and \Hup. This is shown in Fig.\,\ref{Fig5}b. Such behavior can be 
understood 
keeping in mind that the upper critical field in our sample is 
characterized by inhomogeneity broadening. We believe that the sliver 
of magnetic reversibility corresponds to the appearance of 
superconducting islands in our sample. These give 
rise to magnetocaloric effects, but they are isolated within
the bulk and cannot support a screening supercurrent around the 
circumference of the sample, hence the reversible behavior. 
The ``end'' of the peak effect marks the onset of irreversibility
and corresponds to a shoulder in the decreasing field curve. In 
the critical state screening picture, this occurs when continuous 
superconducting paths form around the 
sample and a macroscopic critical current is supported. As the 
field is lowered below $H_{\text{end}}$, flux exit is delayed due to the increase 
in critical current, until past the peak of the peak 
effect, when accelerated flux exit results in a peak in 
$(ds/dH)$ below \Hp. 

\subsubsection{\label{Sec3a4}Fluctuation peak in $(ds/dH)$}
 
In both field-ramp directions, the features due to the peak effect
become less pronounced for higher temperatures, and finally disappear 
even before the previously identified\cite{Park} critical point is
reached. On increasing fields, for $T>7.18\,\text{K}$, a new peak 
appears in $(ds/dH)$ just below the critical field, shown in Fig.\,\ref{Fig5}c. 
This peak feature has already been reported in calorimetric measurements 
in pure Nb and Nb$_{3}$Sn.\cite{Gough,Lortz} It has been attributed to critical 
fluctuations in the superconducting order parameter which set in as the
critical field is approached.\cite{Thouless} In our measurements
we see the effect of critical-fluctuation entropy enhancement in
\dsdh. We expect the same peak to exist in the curves
displaying the peak effect feature, but its presence will be
obscured by the dramatic results of non-equilibrium magnetization
discussed above. Interestingly a similar peak is \textit{not} observed 
on the decreasing field data, Fig.\,\ref{Fig5}c. Moreover, it is evident 
in Fig.\,\ref{Fig5}c, that apart from 
hysteretic behavior over an approximately $100\,\text{Oe}$ wide region 
between \Hh\; and \Hup\, the behavior of the sample is reversible to 
within noise levels. This behavior occurs consistently at all temperatures 
where the peak effect is not observed.

\subsubsection{\label{Sec3a5}The superconducting to normal transition region}

\begin{figure} [!t]
\centering
\includegraphics[scale=0.45] {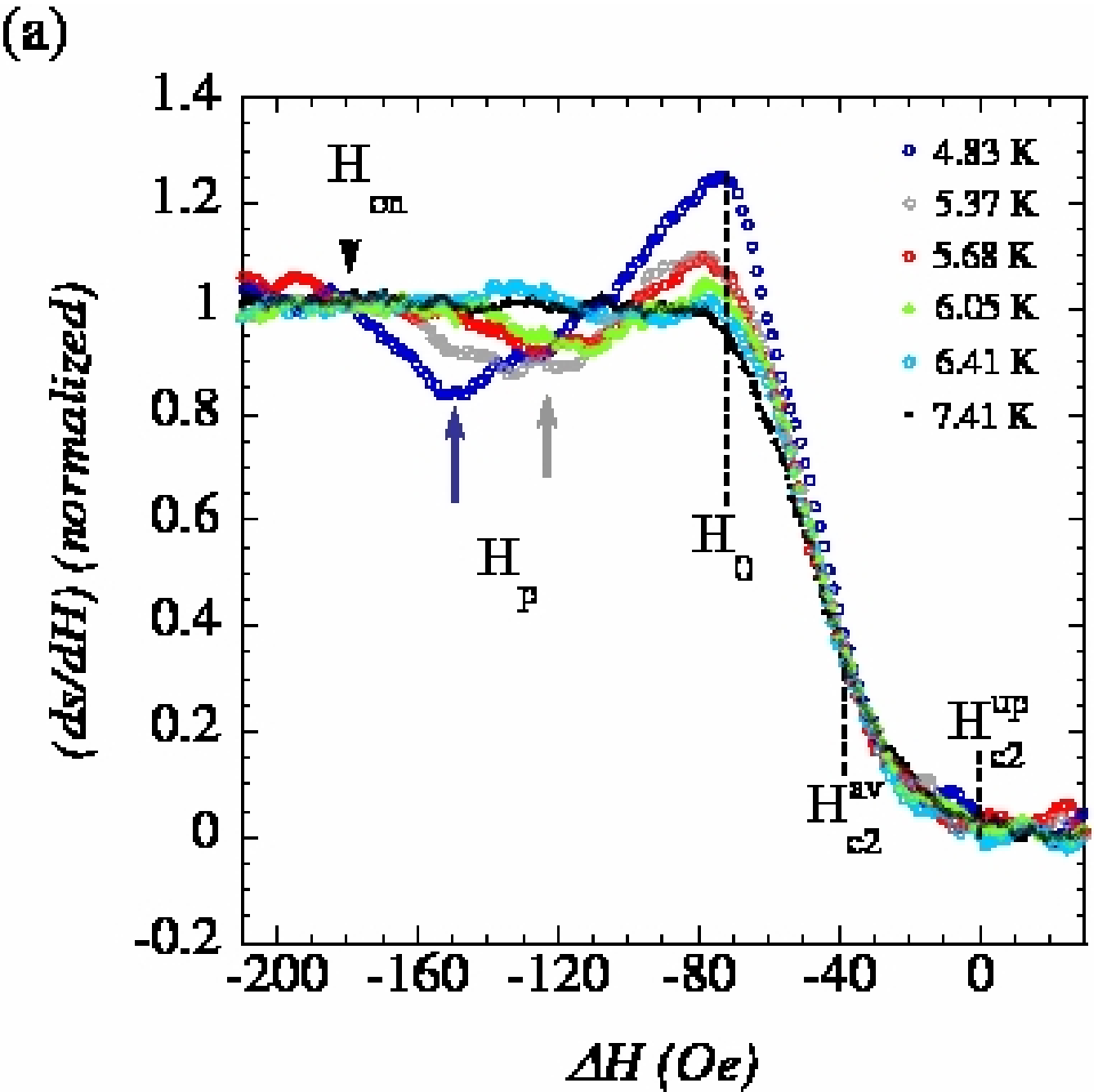}
\vspace*{0.3cm}
\includegraphics[scale=0.45] {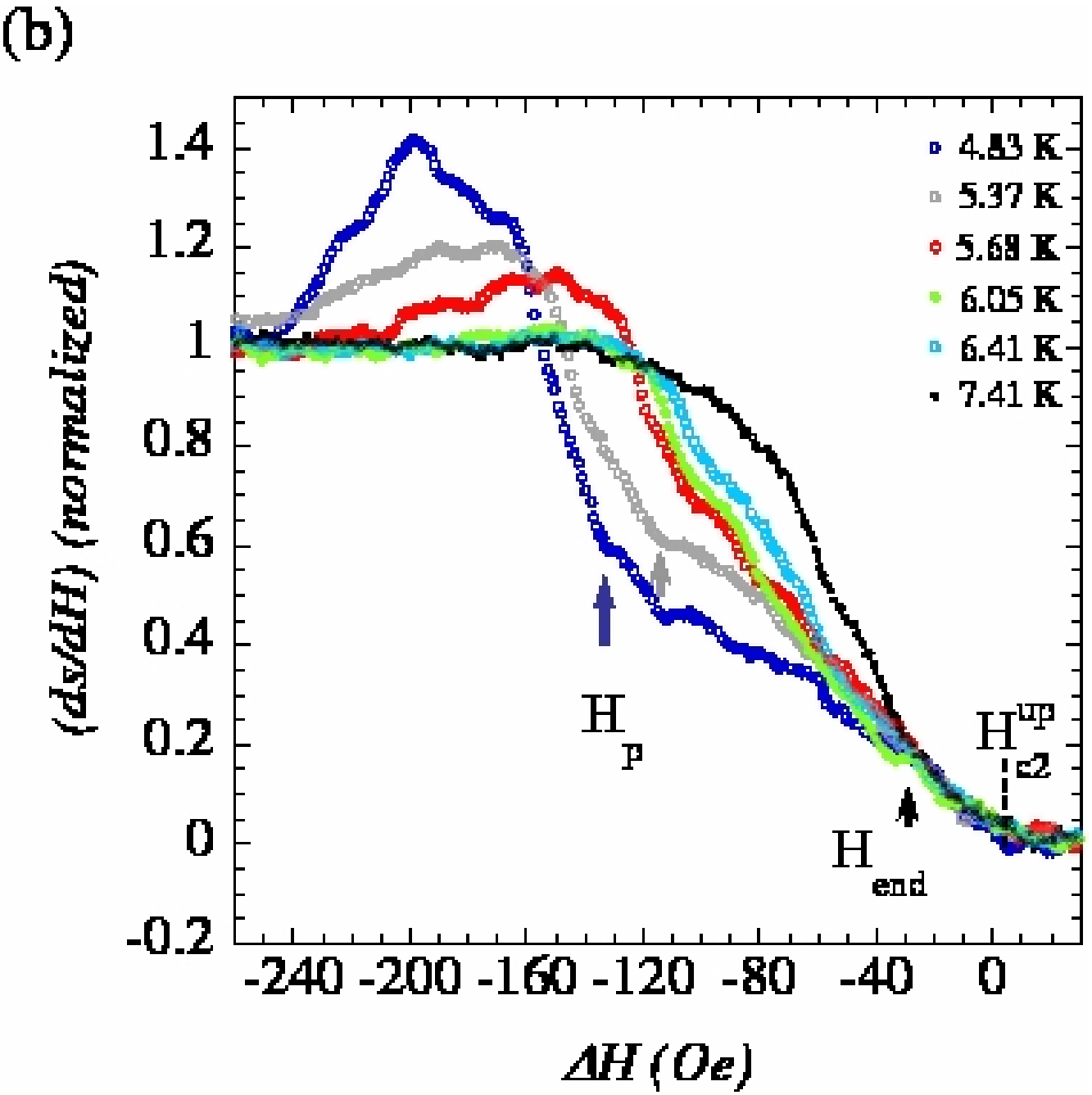}
\vspace*{0.3cm}
\includegraphics[scale=0.45] {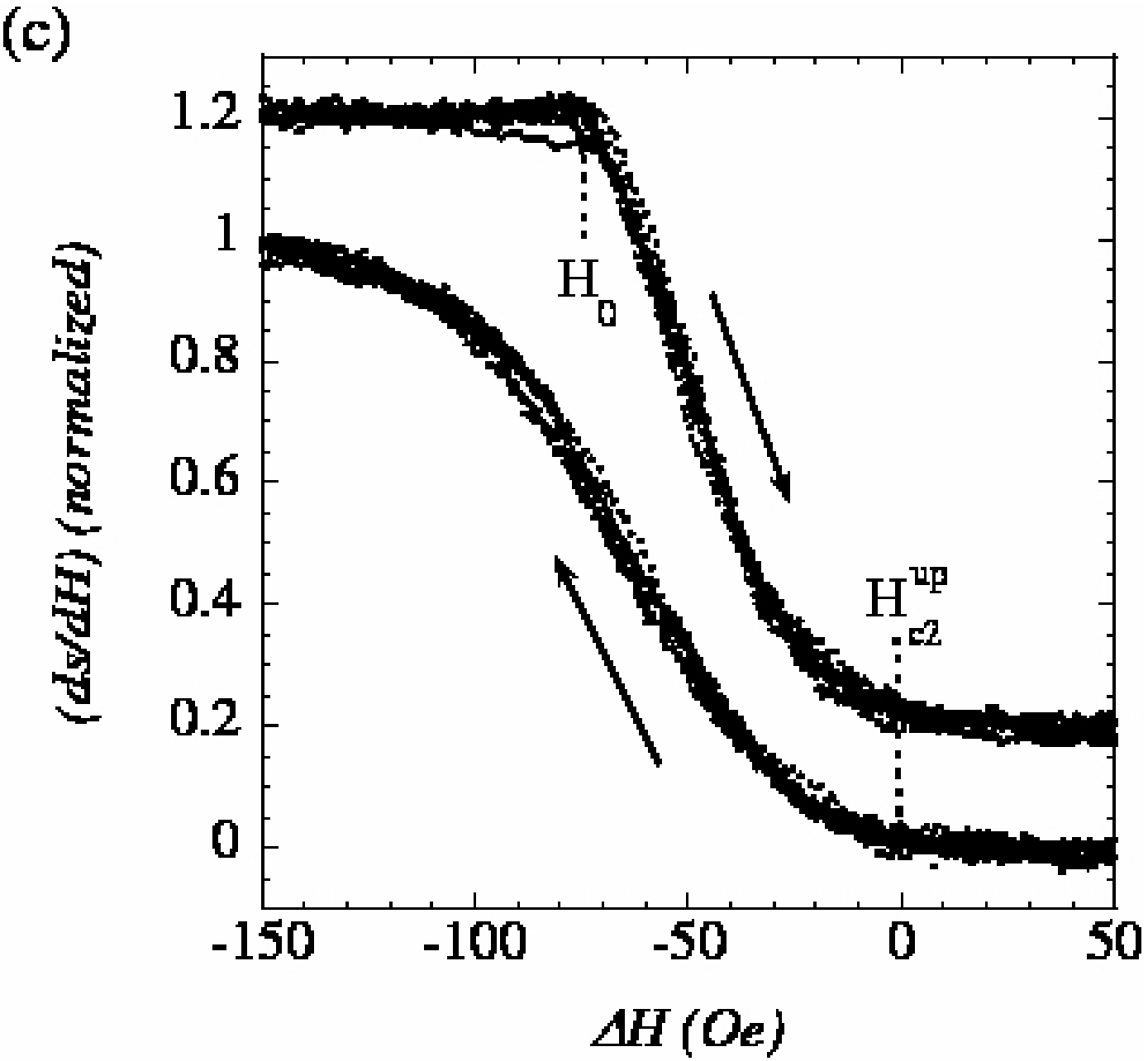}
\vspace*{-0.2cm} \caption{\footnotesize{\textit{(color online)} 
Normalized entropy derivative $(ds/dH)$ versus $\Delta H=H-$\Hup, for the magnetocaloric data
in the upper critical field region. $(ds/dH)$ is rescaled to unity in the $\Delta H$ region between
-300 and -250\,Oe. (a) Increasing field data with and without the peak effect. The 7.41\,K data do not show 
peak effect. (b) Same as a for decreasing field.
(c) Collapsed magnetocaloric curves without peak effect, for 
increasing (top, vertically offset by +0.2) and decreasing (bottom) 
fields. Temperatures 
included: 7.18$\,\text{K}$, 7.41$\,\text{K}$, 7.55$\,\text{K}$, 
7.94$\,\text{K}$, 8.15$\,\text{K}$, 8.33$\,\text{K}$.}} \label{Fig6} 
\vspace*{-0.2cm}
\end{figure}

A very striking feature of our data is the invariance of the 
shape of the transition into (or out of) the bulk normal state
with changing temperature, or critical field. 
For increasing fields, the transition into the normal state 
occurs between the fields \Ho\hspace{0.01cm} and \Hup, where 
$(ds/dH)$ drops to zero monotonically, as shown in Fig.\,\ref{Fig6}a. 
To illustrate this, in Fig.\ref{Fig6} we show the \textit{normalized} 
entropy derivatives as a function of field for temperatures 
ranging from $4.83\, \text{K}$ to $8.33\, \text{K}$, with an expanded 
view of the upper critical field region. The normalization has
been performed such that the average of $(ds/dH)_{norm}$ over a
$50\,\text{Oe}$ wide region below the onset of the peak effect 
equals unity. The curves have also been horizontally offset, on 
a $\Delta H=H-$\Hup\hspace{0.01cm} axis. The horizontal alignment 
can alternatively be performed by aligning the ordinate of either 
\Ho, or the part of the curve where the signal equals a given 
value, for example 0.1 in the normalized Y axis. All different 
criteria result in alignments differing by only a few Oe. The 
case is similar for decreasing fields.

In Fig.\,\ref{Fig6}a we present increasing field data that display 
the peak effect on the normalized/offset axes. For comparison, one 
curve which does not display the peak effect has been included. This 
corresponds to T=7.41\,K. In Fig.\,\ref{Fig6}b we present the corresponding
decreasing field data. In Fig.\,\ref{Fig6}c we show only curves 
without a peak effect, for both increasing and decreasing field. These 
figures illustrate the uniform characteristics of the transition
between the mixed state and the normal state. This is most evident in Fig.\,\ref{Fig6}c: 
All different curves collapse onto one uniform curve for each field ramp direction. 
In figures \ref{Fig6}a\,\&\,b, the occurrence of the peak effect results in variations of the
magnetocaloric signal around this uniform transition. These variations are, 
as already discussed, consistent with critical state induced flux screening
on the field ramps. The uniformity of the transition for all field values implies
that critical fluctuation broadening of the transition plays a minor role in 
our sample. Rather, inhomogeneity broadening seems to be the cause for the observed
behavior.

As already stated in Sec.\,\ref{Sec2}, calorimetric measurements indicate 
inhomogeneity broadening to a width of approximately 83\,mK for the zero 
field transition in our sample. With this in mind, we conclude that for 
increasing fields the gradual disappearance of the magnetocaloric signal 
in the region between 
\Ho\hspace{0.01cm} and \Hup\hspace{0.01cm} corresponds to the gradual loss 
of bulk superconductivity in our sample. In all of our measurements, the 
width $H_{\text{c2}}^{\text{up}}-H_{0}$ is essentially constant around a mean of 
$74.1\pm1.9\, \text{Oe}$ (Fig.\,\ref{Fig6}a\,\&\,c), which translates to a 
width of $78.8\pm2.0\,\text{mK}$ on the temperature axis. This value is in 
good agreement with that obtained in the calorimetric measurement, given the finite 
temperature step of 10 to 15\,mK used in the latter. 
Based on the identification of the lower and upper limits of the upper 
critical field, we identify the location of the mean field transition, 
\Hav, to be in the midpoint of the \Ho\hspace{0.01cm} to \Hup\hspace{0.01cm} 
range, see Fig.\,\ref{Fig6}a. Local variations in electronic properties in 
the sample cause broadening around this value. With this in mind we proceed 
to the discussion of the evolution of the peak effect.

\subsubsection{\label{Sec3a6}Disappearance of the peak effect in $(ds/dH)$}

We already mentioned that in the magnetocaloric measurements 
the peak effect is not observed for temperatures above $7.18\,\text{K}$, 
or critical fields \Hav\hspace{0.01cm} below $1718\,\text{Oe}$. 
The disappearance of the bulk peak effect at such high field 
seems to contradict the previous observation, from ac-susceptometry, 
that the peak effect occurs for fields as low as approximately 
$900\,\text{Oe}$.\cite{Park} Our current findings offer a resolution of 
this controversy. 

\begin{figure} [!tb]
\centering
\includegraphics[scale=0.45] {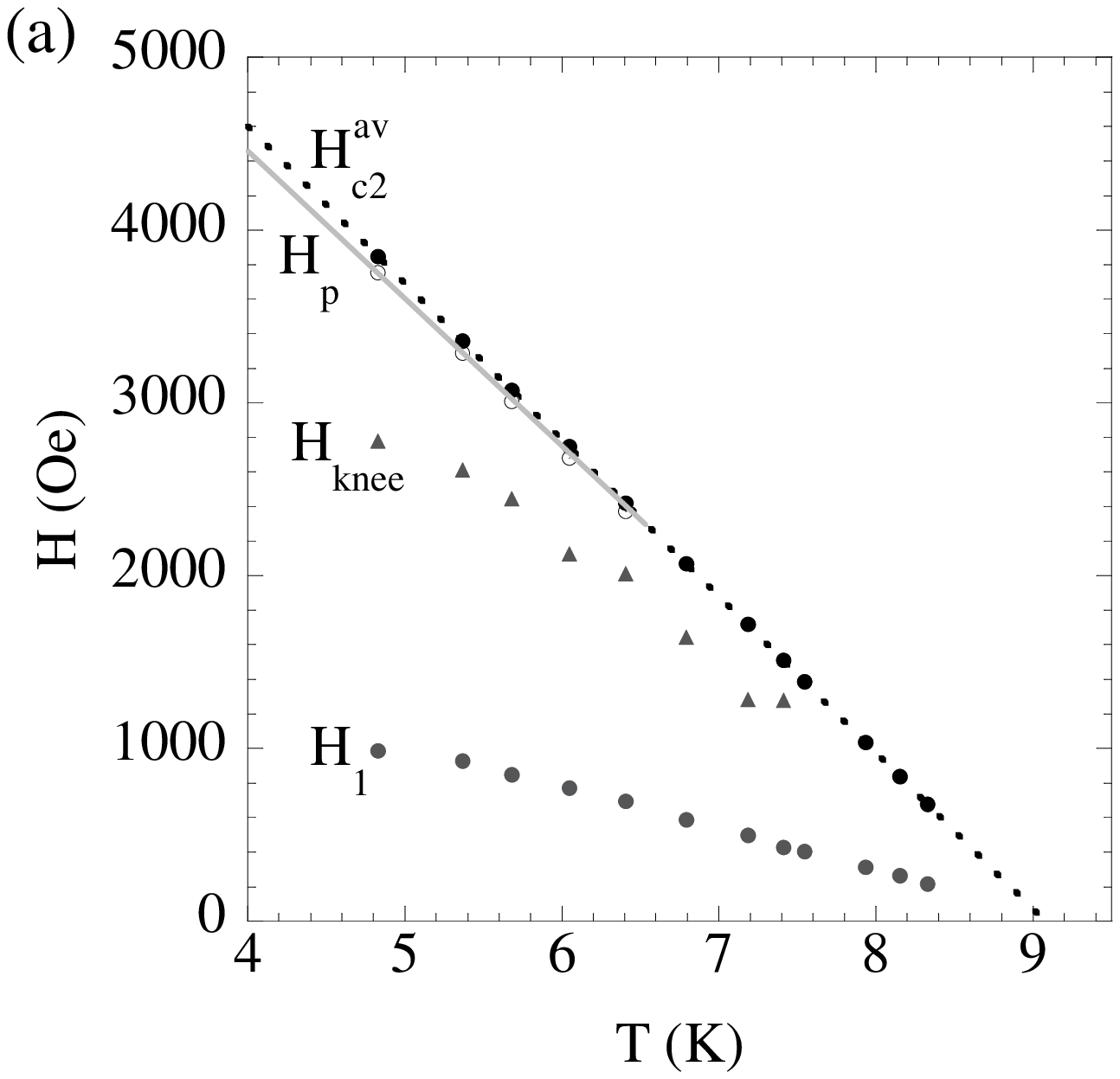}
\includegraphics[scale=0.45] {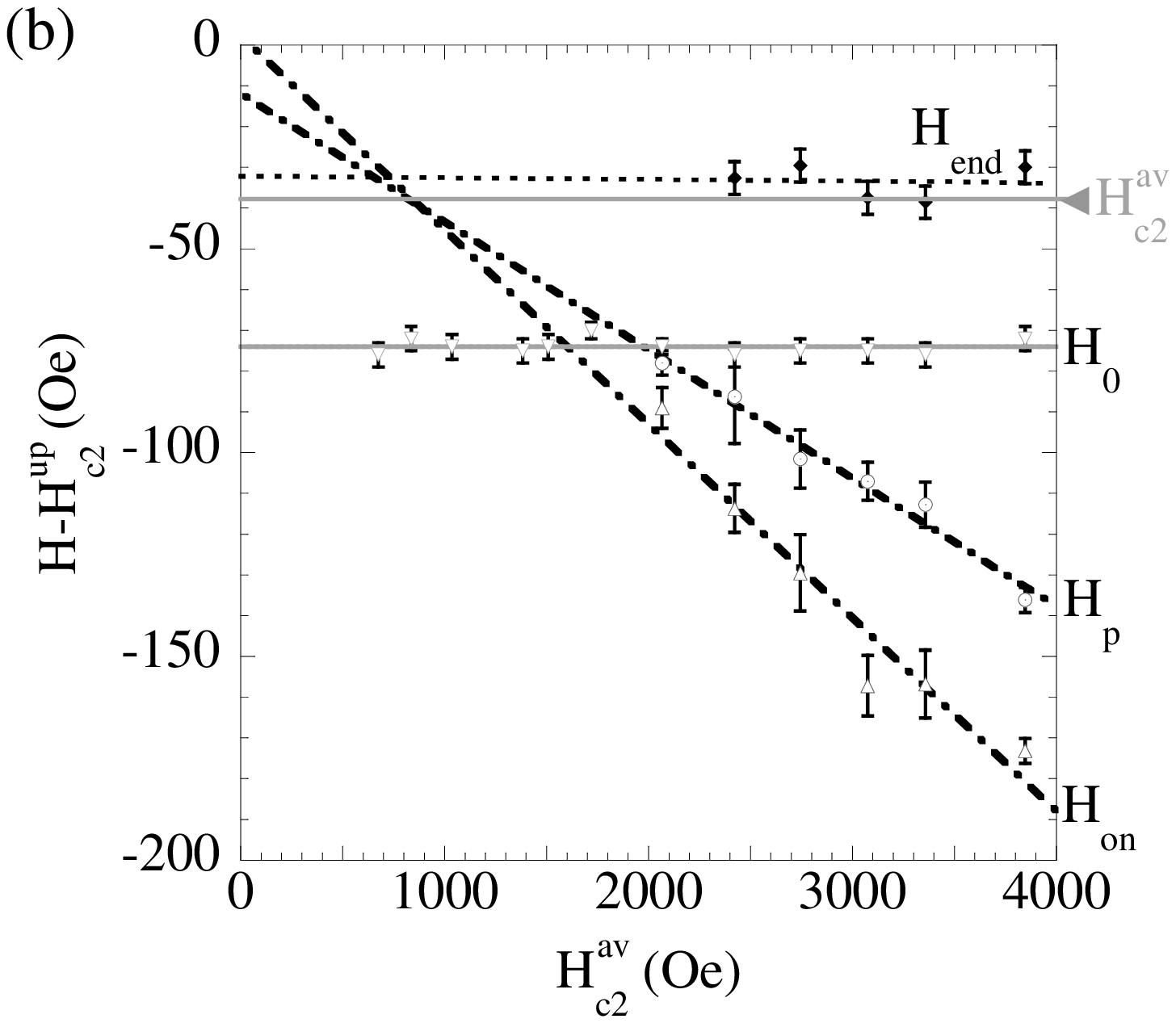}
\vspace*{-0.2cm} \caption{\footnotesize{Phase diagram obtained
from magnetocaloric measurements. (a) The identified features in
$H-T$ axes, and linear fits through the \Hp\, and \Hav\, lines. 
(b) Detail of the peak effect region: distance of peak 
effect features from the conventionally defined \Hup\hspace{0.01cm} 
line, vs. $H_{\text{c2}}^{\text{av}}$.}} 
\label{Fig7} \vspace*{-0.2cm}
\end{figure}

To do this we trace the evolution of the positions of the onset 
and the peak of the peak effect versus \Hav\hspace{0.01cm} in our 
increasing field data.  We focus on increasing fields, because in 
these the range of the upper critical field transition between \Ho\, and \Hup\,
is clearly discernible. The location of each of these features is 
identified by the intersection of two linear fits on different 
sections of the $(ds/dH)$ curve. Each fit is performed in a 
limited $\Delta H$ range on either side of the turning point where 
the feature occurs. For example, \Hp\, is defined by the 
intersection of two linear fits to the data, one roughly in the 
range of \Hon\, to \Hp\, and one in the range of \Hp\, to \Ho. In 
Fig. \ref{Fig7}a we  show the positions of \Hl, $H_{\text{knee}}$,
\Hp, and \Hav\, in $H$ vs. $T$ axes. In 
Fig. \ref{Fig7}b we  mark the positions of the onset, the peak, and 
the end of the peak effect, as well as the lower end (\Ho) and the 
midpoint (\Hav) of the \Hab\hspace{0.01cm} transition in $\Delta H$ vs. 
\Hav\hspace{0.01cm} axes. It is evident in the figure that the peak 
effect disappears when it crosses over into the \Ho\hspace{0.01cm} 
to \Hup\hspace{0.01cm} range, where bulk superconductivity is 
partially lost due to sample inhomogeneity.

It is not clear whether the peak effect continues to exist 
with a reduced magnitude inside this region. It seems likely 
that its magnitude is reduced below detectable levels. This 
can be due to loss of superconductivity in regions of the 
sample and the resulting absence of bulk macroscopic critical 
current. 
However the peak effect continues to manifest itself in 
ac-susceptibility. This can be explained under the assumption 
that the screening supercurrent near the sample surface is 
assisted by the sheath of surface superconductivity. This 
picture also provides an explanation for the approximately 
simultaneous disappearance of the peak effect and surface 
superconductivity in previous studies.\cite{Park} 

The observation with ac-susceptometry of a continuation of the 
peak effect line on the surface of the sample is an indication 
that the peak effect exists, though unobservable, at these lower 
fields, in isolated superconducting islands in the bulk of the 
sample as well. Moreover, as shown in 
Fig. \ref{Fig7}, the linear extrapolation of the onset and the 
peak of the peak effect, as well as \Hav, merge at a field of 
approximately $850\,\text{Oe}$, suggesting that this may indeed 
be the vicinity of the critical point where the first-order phase 
transition underlying the peak effect ends.

\subsubsection{\label{Sec3a7}The \Hab\, and $T_{\text{c2}}$ transition lines}

The nature of the MCP which was previously identified 
by Park \textit{et al.}\cite{Park} remained unresolved. Our current 
findings suggest that surface superconductivity plays only a 
coincidental role in the disappearance of the peak effect. Thus, the 
nature of the MCP is determined from the nature of the \Hab\, and 
$T_{\text{c2}}$ lines discussed in section \ref{Sec1}. 

We have a means of comparing the \Hab\, and $T_{\text{c2}}$ transitions.
The approximate location of the MCP in ac-susceptometry is at $T=$8.1\,K, $H=$900\,Oe. The 
disappearance of the peak effect below $1718\,\text{Oe}$ in the 
magnetocaloric measurements but only below approximately 
$900\,\text{Oe}$ in ac-susceptometry, allows us to compare the 
transitions into the bulk normal state on 
the two sides of the MCP. As shown in Fig.\,\ref{Fig6}c, 
all the curves with \Hav$<1718\,\text{Oe}$ (or equivalently 
$T>7.18\,\text{K}$) for both increasing and decreasing fields 
collapse strikingly on two different curves. These data include 
transitions on both sides of the MCP. This suggests that the 
phase transitions out of the Bragg Glass (\Hab) and disordered  
($T_{\text{c2}}$) phases are of the same nature. In other words, 
well defined vortices exist in the disordered vortex state above 
the peak effect. Here ``well defined'' is taken to mean that their 
magnetocaloric signature is indistinguishable from the one obtained 
in the transition between the normal and the Bragg Glass phases. 
However, it has to be borne in mind that changes in critical behavior 
can be subtle and hard to identify in our sample which shows 
significant inhomogeneity broadening.

\begin{figure}[!b]
\includegraphics[scale=0.45] {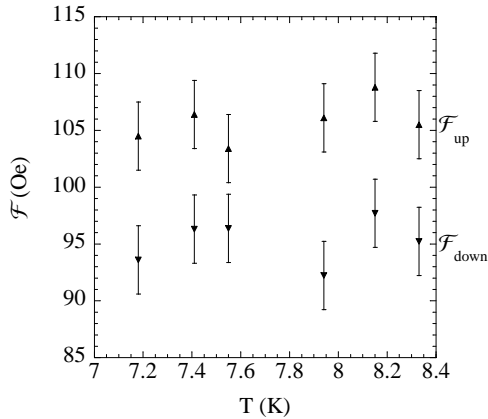}
\caption{Integrals of the rescaled $(ds/dH)$ 
for $\Delta H$ from $-150\,\text{Oe}$ to $30\,\text{Oe}$, for 
increasing ($\mathcal{F}_{up},\;\blacktriangle$) and decreasing 
($\mathcal{F}_{down},\;\blacktriangledown$) fields. The 
peak effect feature was not observed at these temperatures. 
The error bars reflect the uncertainty in alignment of the 
magnetocaloric curves. Temperatures above 8.15\,K correspond 
to the \Hab\, line of transitions proposed by 
Park \textit{et al.}\cite{Park} }
\label{Fig8}
\end{figure}

We can overcome this difficulty and look for changes in critical 
behavior by integrating the normalized experimental curves. This 
way, a change or a trend in critical behavior obscured by 
inhomogeneity broadening will be more easily discerned as a trend 
in the computed integrals. The computed integrals of the normalized 
increasing ($\mathcal{F}_{up}$) and decreasing ($\mathcal{F}_{down}$) field curves 
around the region of the upper critical field are shown 
in Fig.\,\ref{Fig8}. The integration has been performed between 
$\Delta H=-150\, \text{Oe}$ and $30\, \text{Oe}$ in the aligned 
axis. More explicitly:
\begin{equation*}
\mathcal{F}=\int_{\Delta H=-150}^{\Delta H=30}(ds/dH)_{normalized}\,d(\Delta H)\,.
\end{equation*}
The error bars arise from the alignment uncertainty. All 
temperatures refer to the $T_{\text{c2}}$ transition, except for the two
marked with an asterisk which correspond to \Hab. 
We see no systematic trend in the result in any of the available 
temperatures. In conclusion, as far as the inhomogeneity of our 
sample allows us to discern, there is no detectable change between 
the low-field $H_{\text{c2}}$ transition and the high-field 
$T_{\text{c2}}$ transition to the normal state.

\subsubsection{\label{Sec3a8}End of the peak effect}

Finally, we discuss the ``end'' of the peak effect. Its position 
in the phase diagram is shown in Fig.\,\ref{Fig7}b. In our data $H_{\text{end}}$ 
occurs in a range between 32 and 39\,Oe below \Hup\, and slightly above 
the position of the average superconducting transition, \Hav. 
$H_{\text{end}}$ has been identified as the field at which a superconducting 
network which can support a macroscopic screening supercurrent forms inside 
the sample. The ``end'' feature occurs slightly above the midpoint of the 
\Ho\, to \Hup\, range, as shown in Fig.\,\ref{Fig7}b.  This indicates that
the critical current appears when roughly half of the sample is in the mixed 
state while the rest is still in the normal state. This observation is very 
interesting but requires further investigation. The role of surface 
superconductivity in establishing macroscopic supercurrents in the 
superconducting network can be examined experimentally.

\subsection{\label{Sec3b}Discussion}

\subsubsection{\label{Sec3b1}Surface barrier, flux-flow heating, critical state screening}

We can extend the conclusions from our measurements by evaluating the results 
of non-equilibrium and irreversible processes in the magnetocaloric effect. 
 
We start with the surface barrier. Its presence results in delay of flux entry 
into the sample on increasing fields, up to a field approximately equal to
the thermodynamic critical field, $H_{C}$.\cite{deGennes} In addition, the 
surface barrier has the more subtle consequence of introducing an asymmetry 
between the measured \dT\hspace{0.01cm} on increasing and decreasing fields. 
On increasing fields, vortices have to enter the sample through an energy 
barrier in a vortex free region.\cite{Clem, Burlachkov} In this process, 
energy is dissipated approximately at a rate
\begin{equation*}
\Phi_{0}\cdot\big(({H_{C}}^2+H^2)^{1/2}-H\big)\cdot (V/4\pi)\cdot dH/dt\,,
\end{equation*} 
where $V$ is the sample volume and $H_{C}$ the thermodynamic critical field. 

\begin{figure} [ht]
\centering
\includegraphics[scale=0.43] {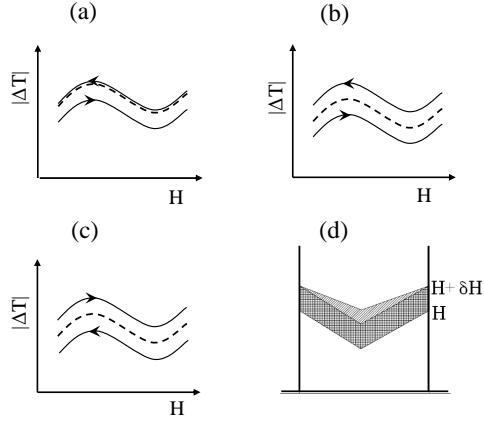}
\caption{\footnotesize{ (a)-(c) Schematic representations of  the asymmetries induced by 
irreversible and non-equilibrium effects on the magnetocaloric measurement, 
$\left|\Delta T\right|$. The field ramp directions are indicated by arrows, and 
the equilibrium \dsdh\, curve corresponds to the dashed line. (a) Surface barrier 
heating. (b) Flux flow heating. (c) Critical state screening.\\ (d) Qualitative schematic 
of the evolution of the critical state field profile, for increasing 
field and $\partial j_{c}/\partial H\,<\,0$. When the applied field increases from $H$
to $H+\delta H$, the flux increase corresponds to the total shaded area, not just the lower
shaded part. This results in increased magnetocaloric signal, as shown in (c) and discussed in 
the text.}}
\label{Fig9}
\end{figure}

In our measurements, for example at $H\approx3000\, \text{Oe}$,
this amounts to approximately $2\, \mu \text{W}$, 
and will reduce the (negative) \dT\hspace{0.01cm} 
observed on increasing fields by roughly $1\, \text{mK}$. On decreasing fields, 
the surface barrier has essentially no effect, and no irreversible 
heating is expected.\cite{Clem} An asymmetry of this kind is shown schematically
in Fig.\,\ref{Fig9}a, and it is present in our data, for example in Fig.\,\ref{Fig5}b. 
Moreover, this type of asymmetry disappears for lower values of the
critical field, where surface superconductivity has disappeared,\cite{Park} for 
example in Fig.\,\ref{Fig5}c.

Flux flow heating of the sample also leads to a similar asymmetry between the 
ascending and descending field branches. On increasing fields the negative 
\dT\hspace{0.01cm} is reduced and on decreasing fields the positive 
\dT\hspace{0.01cm} increased. An order of magnitude estimate of flux-flow 
heating can be obtained on the basis of the Bardeen-Stephen model.\cite{BardSteph} 
For a cylindrical sample of radius $R$, length $L$, and for smooth field 
ramping at a rate $dH/dt$, one obtains 
$P_{ff}\approx10^{7}(dH/dt)^2\pi R^4 L/(8\rho_{ff})$. This turns out to be 
negligible for the field ramp rates, approximately 1\,Oe/sec, used in our 
measurements. We show the effect of this mechanism, 
grossly exaggerated, in Fig.\,\ref{Fig9}b. Between the above 
two sources of irreversible heating, it is clear that low field ramp 
rates will render the latter ($\propto (dH/dt)^{2}$) negligible, 
but will not reduce the effect of the former which scales as $dH/dt$, as 
does the magnetocaloric \dT. 

Next we examine the case of a non-equilibrium critical state profile 
outside the peak effect region. A critical current that monotonically 
decreases with field, i.e. away 
from the peak effect region, will result in the opposite asymmetry than the 
one just mentioned. This is so, because on increasing fields the critical 
state profile becomes less steep, resulting in faster loss of flux than 
the field ramp rate, and thus increased \dT. The opposite occurs on 
decreasing fields, resulting in a lower \dT\hspace{0.01cm} than indicated 
by the equilibrium \mt. This mechanism is shown schematically in Fig.\,\ref{Fig9}d.
A simplified calculation for cylindrical sample 
geometry, like the ones found in the literature, \cite{Bean,Xu} yields 
an asymmetry factor in \dT\hspace{0.01cm} equal to: 
\begin{equation} \label{Eq3}
			1\pm(4\pi R/3c)(\partial J_{c}/\partial H)_{T}
\end{equation}
\noindent
for increasing ($-$) and decreasing ($+$) fields. Here, $R$ is the sample 
radius and $J_{c}(H)$ the $H$ dependent critical current. Moreover it is 
assumed that the critical state field profile does not result in significant 
modification of the critical current in the sample interior. Nevertheless, 
even in the case where the value of the critical current varies spatially 
inside the sample, $J_{c}=J_{c}(B(r))$, the same type of asymmetry results, 
with slightly modified value though. The derivation of the above factor 
is outlined in the appendix. This asymmetry is illustrated in Fig.\,\ref{Fig9}c. 
This type of asymmetry is only observed in our data in the region of the 
upper critical field and the peak effect. 

For example, in data without peak effect, as shown in Fig.\,\ref{Fig5}c, 
the magnetocaloric curves are  reversible outside a region between 
\Hh\; and \Hup\, but consistently show hysteresis in that region, see 
also Fig\,\ref{Fig6}b. This behavior can be attributed to critical state 
screening. Then, the hysteresis which persists almost all the way up to \Hup\, 
indicates that the critical current is nonzero arbitrarily close to the 
upper critical field, and vanishes abruptly only at a distance of $10$
-$30\,\text{Oe}$ from \Hup. In addition, the lack of asymmetry in the 
increasing and decreasing field curves below the hysteretic region, 
implies that the critical current density varies very weakly with applied 
field. More specifically, the asymmetry in any of the data without 
peak effect is less than 5\% of the magnitude of $(ds/dH)$. 
By Eq.\,\ref{Eq3} we obtain
\begin{equation*}
(\partial J_{c}/\partial H)_{T}\,<\,(1/2)\cdot0.05\cdot(3c/4\pi R)\approx\,0.36\,\text{A}\,/\,\text{cm}^{2}\,\text{Oe}\,,
\end{equation*} 
where the numerical result is expressed in conventional units for
convenience. 

\subsubsection{\label{Sec3b2}Critical current estimates}

From the measurements we can estimate the value of the critical current 
before it vanishes in the upper critical field region, in the context 
of critical state screening. We assume that the entropy per vortex is 
essentially constant in the neighborhood of the upper critical 
field and the peak effect.\cite{note} Therefore changes in magnetic flux ($\Delta\Phi$) 
are proportional to changes in entropy, $\Delta\Phi\propto\Delta S$. Then an 
integral of $ds/dH$, such as $\mathcal{F}_{up}$ and $\mathcal{F}_{down}$, can be approximately 
taken to represent a change in magnetic flux in the sample. In the critical
state model, we can relate changes in flux to the critical current.

\begin{figure}
\includegraphics[scale=0.45] {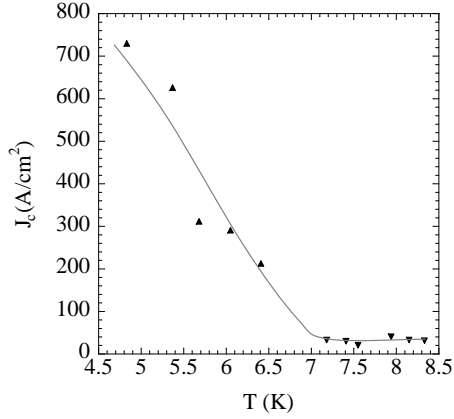}
\caption{\footnotesize{Estimated values of critical current in the neighborhood 
of the upper critical field in our sample. For T$\,>7\,$K we show $J_{c}$ (marked by 
$\blacktriangledown$), the critical current at the limit of hysteresis, \Hh, in 
temperatures where no peak effect was observed. At temperatures below 7.18\,K 
where peak effect was observed we show $J_{c}^{p}$ ($\blacktriangle$), the estimated 
critical current at the peak of the peak effect. The line is a guide to the eye.}}
\label{Fig10}
\end{figure}

At the temperatures where no peak effect is observed, we estimate 
the critical current in the neighborhood of \Hh. For simplicity we 
consider a linear field profile in the sample, with slope
\begin{equation*} 
dH/dr=(4\pi/c)\,J_{c}\,, 
\end{equation*}
where $r$ is the radial distance from the axis 
of the (cylindrical) sample. In the vicinity of \Hh\, simple 
integration gives a macroscopic magnetic flux difference ($\Delta\Phi$) 
between the increasing and decreasing field branch approximately equal to: 
\begin{equation*}
\Delta \Phi=\left(1+\frac{1}{\zeta}\right)(16\pi^{2}/3c)\,J_{c}(H_{\text{hyst}})\,R^{3}\,.
\end{equation*} 
see Eq.\,\ref{EqA2} in the appendix. Due to the normalization performed for 
the integrals in Fig.\,\ref{Fig8}\, the flux difference is also approximately given by: 
\begin{equation*}
\Delta \Phi=(\mathcal{F}_{up}-\mathcal{F}_{down})\,\pi R^{2}\,.
\end{equation*} 
These two relations, allow us to obtain  estimates for the critical current 
in the sample before this collapses to zero in the upper critical field 
region. We show these in Fig.\,\ref{Fig10}, for temperatures above 7\,K, 
corresponding to the symbol for $J_{c}$. One should note that critical 
currents with the values given for $J_{c}$ in the figure, will lead to 
radial variation of the induction inside the sample by roughly 10\,Oe. 
Due to the constraint imposed on $(\partial  J_{c}/\partial H)_{T}$, 
(see discussion above), the critical current density will vary inside the 
sample due to screened induction by less than 3.6\,A/cm$^{2}$. 
This represents only 10\% of its value, and thus the assumption that 
the slope of the critical state profile (i.e. $J_{c}(r)$) is essentially 
constant is self consistent.

A similar procedure can be followed for the curves showing the peak 
effect, in order to estimate the critical current at the peak of 
the peak effect. This is most easily done for the decreasing field, 
curves, under the additional assumption that the critical current 
increases linearly from zero at $H_{\text{end}}$ to $J_{c}^{p}$ at 
\Hp. The integration relating the screened flux ($\Delta\Phi$) to 
these three quantities leads to the rather complicated equation:
\begin{equation}\label{Eq4}
			\left\{ \Delta H-\frac{\Delta\Phi}{\pi R^2(1+\frac{1} {\zeta})} \right\}x^{2}-2\Delta Hx+2\Delta H=2\Delta H e^{-x}\;,
\end{equation} 
where $\Delta H =H_{\text{end}}-$\Hp, $x=(4\pi\,R\,J_{c}^{p})/(c\,\Delta H)$. 
The screened flux for a curve with  peak effect is obtained via integration
between \Hp\, and \Hup with respect to the universal curve of 
Fig.\,\ref{Fig6}b (bottom). The resulting equations are solved numerically, 
and the results for $J_{c}^{p}$ are shown in Fig.\,\ref{Fig10}. The value 
at $6.79\,\text{K}$ is not included, because the peak effect is barely 
observable at that temperature, which makes the procedure we outlined inapplicable. 

\subsubsection{\label{Sec3b3}Low field hysteresis: alternative interpretaion}

We return to the low-field hysteresis near \Hab, illustrated in Fig.\,\ref{Fig5}b. 
The combination of hysteretic and reversible behavior in the vicinity of \Hab\;
is striking. Furthermore this behavior persists in all of 
the measurements which do not show the peak effect, down to an upper critical 
field of $676\,\text{Oe}$. In addition, we suggested earlier that the 
disappearance of the peak effect in our bulk measurements is due to loss of 
superconductivity in parts of the sample, while in ac-susceptometry it is due to the 
disappearance of surface superconductivity. Thus the possibility arises that 
the peak effect continues to exist at low fields, but it is not observable 
with the techniques used so far. In other words, our data raise the question 
whether the hysteretic order-disorder transition reported previously\cite{Ling} 
continues down to low fields, where no peak effect is observed, but hysteresis 
occurs over a range too narrow to be detectable in neutron scattering. In this 
scenario, the hysteresis seen in Fig.\,\ref{Fig5}c is related to a first order 
phase transition. If this is the case the hysteresis presented here will be 
observable in high resolution calorimetric measurements. 

\subsubsection{\label{Sec3b4}The ``knee'' feature}

Finally we return to the newly identified ``knee'' feature, shown in Fig.\,\ref{Fig4}c. 
From the occurrence of the knee in both ramp directions we conclude that it 
corresponds to an equilibrium feature of the thermodynamic behavior inside 
the Bragg Glass phase. To appreciate this argument, note that all three 
sources of irreversible and non-equilibrium effects discussed at the 
beginning of this subsection, will induce asymmetry on the $(ds/dH)$ curves 
for opposite field-ramp directions. For example, the two curves of 
Fig.\,\ref{Fig4}c show asymmetry, as already mentioned. This is due to 
surface barrier-related heating on increasing field. On the other hand, 
symmetric trends in the measured $(ds/dH)$ curves have to be related to 
equilibrium behavior, and we thus conclude that $H_{\text{knee}}$ 
corresponds to an equilibrium feature. Neutron scattering did not reveal 
any structural changes in the vortex lattice\cite{Ling,Park} around 
$H_{\text{knee}}$, which suggests that the nature of this feature is 
rather subtle. It would be interesting to investigate the corresponding 
part of the phase diagram for changes in dynamical response, as well as 
for a possible relation of this novel feature to the thermomagnetic 
instability in Nb.

\subsection{\label{Sec3c}Summary}

From the reported magnetocaloric-effect results, we gain a 
significant amount of novel information about the Nb sample 
previously studied using SANS and ac-susceptometry. 

The upper critical field shows significant 
inhomogeneity broadening. The broad \Hab\hspace{0.01cm} 
transition is related to the disappearance of the peak effect 
in bulk measurements. The peak effect disappears in the 
phase diagram when it enters the range where regions of the 
sample are normal. Yet, magnetic irreversibility, indicating
nonzero critical currents, occurs when regions of the sample are
normal. The peak effect is observable with ac-susceptometry at 
fields much lower than with magnetocaloric measurements. In the 
former case screening occurs locally, near the surface,\cite{Park} 
presumably assisted by a superconducting surface layer. This could 
indicate that the peak effect still occurs in the bulk of the sample
in regions that are superconducting. 

The low field transition from the Bragg Glass phase into the normal 
state seems to be the same as the high field transition between 
the structurally disordered\cite{Ling} vortex state and the 
normal state. Moreover, hysteresis occurs in magnetocaloric 
curves that do not display the peak effect, but in the 
neighborhood of this hysteresis the non-equilibrium and 
irreversible effect signatures that we discussed are absent. 
Finally a new feature corresponding to a knee 
in the magnetocaloric coefficient has been identified and its 
position mapped out in the phase diagram, Fig. \ref{Fig7}\,a. 
These new findings allow us to refine the previously proposed 
peak effect phase diagram (Fig.\,\ref{Fig1}), but also point 
to an alternative picture.

In the multicritical point picture, the magnetocaloric-effect 
results reported here have strong implications for understanding 
the nature of the multicritical point where the peak effect 
disappears.\cite{Park,Adesso,Jaiswal} A tricritical point can be 
ruled out since the change in slope between the \Hp\hspace{0.01cm} 
and low field \Hab\hspace{0.01cm} lines would lead to violation 
of the 180$^\circ$ rule\cite{Wheeler,Park} imposed by thermodynamics. 
The fact that the magnetocaloric transition appears to be 
continuous and of the same character across both $T_{c2}$ and 
$H_{c2}$, suggests that the critical point is 
bicritical. Bicriticality implies the competition of two bulk 
vortex phases, an ordered Bragg glass\cite{natt,GL} and a 
disordered vortex glass. The vortex glass phase is not necessarily 
superconducting in the sense of the original 
proposal,\cite{FisherI} but it has to be a genuine phase possessing 
an order parameter absent in the normal state,\cite{Menon} and 
\textit{in competition} with that of the Bragg glass. In addition, it
possesses superconducting phase rigidity even under partial loss of 
bulk (mixed state) superconductivity.

Finally, we wish to discuss an alternative picture in which the order-disorder 
transition does not end at the previously identified critical point.
The experimental disappearance of the peak effect in magnetocaloric 
measurements may be due to sample inhomogeneity. In addition, the peak 
effect disappears in ac-susceptometry due to the disappearance of 
surface superconductivity.\cite{Park} Nevertheless, the vortex
lattice disordering transition can continue to lower fields, where the 
peak effect is not detectable. Hysteresis related to a first order 
phase transition occurs over a narrow range in this part of the phase
diagram, and the width of the hysteresis region grows appreciably only 
at higher fields where the peak effect is sufficiently far from the 
\Hab\hspace{0.01cm} region. In this picture, no critical point 
associated with the disordering transition exists at any finite field 
or temperature.\cite{Menon} 

\section{\label{Sec4}Conclusions}

We have identified the magnetocaloric signatures of the peak effect in a Nb 
single crystal. In addition we classified and outlined the various sources of 
irreversible and non-equilibrium effects occurring in such measurements 
and offered ways for their evaluation. With these in mind, magnetocaloric
measurements prove very useful in studies of the mixed state. They are
suitable for studying the upper critical field region. They allow the 
identification and study of dynamical effects, such as the onset, the peak 
and the end of the peak effect, and provide estimates of the critical 
currents involved. Moreover they are useful in identification 
of changes in thermodynamic behavior deeper inside the mixed state, e.g. 
the ``knee'' feature identified here. 

With the wealth of novel information obtained in our measurements we were 
able to shed light into the diasppearance of the peak effect in Nb. 
We refined the multicritical point scenario drawn previously.\cite{Park}
In addition, based on experimental facts, we proposed an alternative scenario 
which can be experimentally tested.
\\\\
We thank J. J. Rush and J. W. Lynn from NIST for providing the Nb crystal and 
acknowledge numerous helpful discussions with D. A. Huse, J. M.
Kosterlitz and M. C. Marchetti.  This work was supported by NSF
under grant No. NSF-DMR 0406626.
\\\\

\appendix*
\section{Magnetic flux changes due to the critical state}

\subsection{\label{A1}Asymmetry factor of $(ds/dH)$}

First we derive the critical state asymmetry factor of $(ds/dH)$ discussed 
in Section \ref{Sec3b}. We limit our discussion to cylindrical sample geometry, 
with the field applied along the cylindrical axis.
The critical current is taken to depend on $B$, or 
equivalently on $H$. We focus on  the region close to \Hab, where these are 
linearly related by : $B=(1+1/\zeta)H-H_{c2}/\zeta$, with 
$\zeta=\beta_{A}\,(2\kappa^{2}-1)$.\cite{deGennes2} Due to critical current 
screening, the local induction (or field) inside the 
sample is modified with respect to the applied field and depends on radial 
distance from the axis ($r$) of the sample: $H=H(r)$. From the Amp{\`e}re law 
the field variation resulting from the critical current is: 
\begin{equation*}
			\frac{dH}{dr}=\pm\frac{4\pi} {c}\,J_{c}\big(H(r)\big)\;, 
\end{equation*}
\noindent
for increasing ($+$) and decreasing ($-$) field. 
The critical current depends on $B$, but we make the simplifying assumption 
that for a given value of applied field $H_{a}$, the local induction inside 
the sample leads to a negligible critical current variation: $J_{c}(r)=J_{c}=const$. 
In addition we neglect demagnetizing effects. We then obtain:
\begin{equation*}
			H(r)=H_{a}\pm(4\pi/c)\,J_{c}\cdot(r-R)\;,
\end{equation*}
\noindent 
where $R$ is the sample radius  and again the solutions correspond to increasing ($+$) and
decreasing ($-$) field. The corresponding magnetic induction is simply: 
\begin{equation*}
			B(r)=\left\{H_{a}\pm\frac{4\pi}{c}\,J_{c}\cdot(r-R)\right\}\cdot(1+1/\zeta)-H_{c2}/\zeta\;. 
\end{equation*}
\noindent
This is easily integrated over the cross sectional area of the sample, in order to obtain 
the magnetic flux through the sample:
\begin{equation}\label{EqA1}
			\Phi=\int_{0}^{R}B(r)\,2\pi r\cdot dr\;,
\end{equation}
to yield:
\begin{equation}\label{EqA2}
			\Phi=\left\{H_{a}\,\left(1+\frac{1}{\zeta}\right)-H_{c2}/\zeta\mp\left(1+\frac{1}{\zeta}\right)\,\frac{4\pi R\,J_{c}}{3c}\right\}\cdot\pi
			R^{2}\;.
\end{equation}
\noindent Note that here the significance of the signs is reversed for increasing 
($-$) and decreasing ($+$) field. The rate of change in magnetic flux through the sample 
for changes in applied field is:
\begin{equation*}
				\frac{d\Phi}{dH_{a}}=\left\{1\mp\frac{4\pi R}{3c}\,\left(\frac{\partial J_{c}}{\partial H}\right)_{T}\right\}\cdot\left( 
				1+\frac{1} {\zeta} \right)\cdot\pi R^{2}\;.
\end{equation*}
\noindent This is proportional to the magnetocaloric signal, and includes the 
asymmetry factor given in Eq.\,\ref{Eq3}, with negative ($-$) sign for increasing 
field, positive ($+$) for decreasing. 

\subsection{\label{A2} Flux screening in the peak effect region} 

The critical current between \Hp\; and $H_{\text{end}}$ is taken to be:
\begin{equation*}
  J_{c} = 
  	\begin{cases}
  		0\,, & \text{if $H>H_{\text{end}}$} \\
  		J_{c}^{p}\left(\frac{H_{\text{end}}-H}{H_{\text{end}}-H{\text{p}}}\right)\, , & \text{if 									     										$H_{\text{p}}<H<H_{\text{end}}$}
  	\end{cases}
\end{equation*}
\noindent Then the Amp{\`e}re law for the field ($H_{\text{p}}<H<H_{\text{end}}$) 
inside the sample is:
\begin{equation*} 
			\frac{dH}{dr}=-\frac{4\pi\,J_{c}^{p}}{c}\cdot\left(\frac {H_{\text{end}}-H}{H_{\text{end}}-H_{\text{p}}}\right)
\end{equation*}
\noindent This is solved for applied field $H_{a}=H_{\text{p}}$, with 
the boundary condition $H(r=R)=H_{\text{p}}$ (again by neglecting demagnetizing effects), and yields:
\begin{equation*}
			H(r)=H_{\text{end}}-(H_{\text{end}}-H_{\text{p}})\,e^{(r-R)/l}\;,
\end{equation*}
with 
$l^{-1}=(4\pi\,J_{c}^{p})/(c\,(H_{\text{end}}-H_{\text{p}}))$. From this, the  
local induction is obtained, and via integration over the sample 
cross-section, see Eq.\,\ref{EqA1} the magnetic flux through the sample is:
\begin{eqnarray*}
			\Phi=&&\left( 1+\frac {1}{\zeta} \right)\nonumber\\
					 &&\times\left\{ H_{\text{end}}\,\pi R^{2}-2\Delta H\pi(Rl-l^2)-
			     2\Delta H\pi l^2e^{-R/l}\right\}\nonumber\\
					 &&-H_{c2}\,\pi R^2/\zeta\,,
\end{eqnarray*}
\noindent where we used the notation $\Delta H=H_{\text{end}}-H_{\text{p}}$. The flux
in the absence of screening ($J_{c}=0$) is:
\begin{equation*}
			\Phi'=\left( 1+\frac{1}{\zeta} \right)\,H_{\text{p}}\,\pi R^2-H_{c2}\,\pi R^2/\zeta\,.
\end{equation*}
\noindent Therefore the amount of screened flux $\Delta\Phi=\Phi-\Phi'$, turns out to be:
\begin{eqnarray*}
			\Delta\Phi=&&\pi R^2\,\left( 1+\frac {1}{\zeta} \right)\nonumber\\
								 &&\times\left\{\Delta H-2\,\Delta H\left(\frac{l}{R}-\frac{l^2}{R^2}\right)
									-2\,\Delta H\frac{l^2}{R^2}\,e^{-R/l}\right\}\,.
\end{eqnarray*}
\noindent The additional definition $x=R/l$, straightforwardly leads to Eq.\ref{Eq4}:
\begin{equation*}
			\left\{ \Delta H-\frac{\Delta\Phi}{\pi R^2\left( 1+\frac {1}{\zeta} \right)} \right\}x^{2}-2\Delta H x+2\Delta H=2\Delta H e^{-x}\,.
\end{equation*} 
\\\\\\\\

\bibliographystyle{unsrt}

\begin{thebibliography}{99}

\bibitem{IM}
A. I. Larkin, Sov. Phys. JETP {\bf 31}, 784 (1970); Y. Imry and S.
Ma, Phys. Rev. Lett. {\bf 35} 1399 (1975).
\bibitem{natt}
T. Nattermann, Phys. Rev. Lett., {\bf 64}, 2454 (1990).
\bibitem{GL}
T. Giamarchi and P. Le Doussal, Phys. Rev. Lett., {\bf 72}, 1530, (1994).
\bibitem{Christen}
D. Christen, F. Tasset, S. Spooner, and H. A. Mook, Phys. Rev. B {\bf 15}, 4506 (1977).
\bibitem{Ling}
X. S. Ling, S. R. Park, B. A. McClain, S. M. Choi,, D. C. Dender and J. W. Lynn ,  Phys. Rev. Lett. {\bf 86}, 712 (2001).
\bibitem{Troy}
A. M. Troyanovski, M. van Hecke, N. Saha, J. Aarts, and P. H. Kes,  Phys. Rev. Lett. {\bf 89},
147006 (2002).
\bibitem{PE}W. DeSorbo, Rev. Mod. Phys. {\bf 36}, 90 (1964).
\bibitem{Park}
S. R. Park, S. M. Choi, D. C. Dender, J. W. Lynn,  and X. S. Ling, , Phys. Rev. Lett. {\bf 91}, 167003 (2003)
\bibitem{Adesso}
 M. G. Adesso, D. Uglietti, R. Flukiger, M. Polichetti, and S. Pace, Phys. Rev. B {\bf 73}, 092513 (2006).
\bibitem{Jaiswal}
 D. Jaiswal-Nagar, D. Pal, M. R. Eskildsen, P. C. Canfield, H. Takeya, S. Ramakrishnan, and A. K. Grover, PRAMANA-J. of Phys. {\bf 66}, 113 (2006).
\bibitem{SurfMelt}
R. L. Cormia, J. D. Mackenzie, J. D. Turnbull, J. Phys. Chem., {\bf65}, 2239 (1963);
J. Dages, H. Gleiter, J. H. Perepezko, Mat. Res. Soc., \textit{Symposium Proceedings}, {\bf57} (1986);
R. W. Cahn, Nature, {\bf323}, 667 (1986).
\bibitem{Lortz} 
R. Lortz, F. Lin, N. Musolino, Y. Wang, A. Junod, B. Rosenstein, and N. Toyota, Phys. Rev. B, {\bf74} 104502 (2006).
\bibitem{Daniilidis}
N. Daniilidis \textit{et al.} \textit{(unpublished)}.
\bibitem{Shapira}
Y. Shapira and L.J. Neuringer, Phys. Rev. {\bf154}, 375 (1967).
\bibitem{Dimitrov}
I. K. Dimitrov, \textit{Ph.D. Thesis}, Brown University (2007).
\bibitem{Thermomagnetic}
A. T. Fiory and B. Serin, Phys. Rev. Letters {\bf 16}, 308 (1966);
F. A. Otter Jr., and  P. R. Solomon, Phys. Rev. Letters {\bf 16},
681 (1966).
\bibitem{adiabatic}
In perfectly adiabatic conditions, the heat is ``exchanged'' within the superconductor itself resulting in
a change of its temperature. 
\bibitem{BeanLiv}
C. P. Bean and J. D. Livingston, Phys. Rev. Lett. {\bf 12}, 14 (1964).
\bibitem{FF}
J. Bardeen and M. J. Stephen, Phys. Rev. {\bf 140}, A1197 (1965).
\bibitem{Bean}
C. P. Bean, Phys. Rev. Lett. {\bf 8}, 250 (1962).
\bibitem{deGennes}
P. G. deGennes, Solid State Commun. {\bf 3}, 127 (1965).
\bibitem{Clem}
J. R. Clem, \textit{Low Temperature Physics-LT13}, edited by K. D. Timmerhaus, W. J. O'Sullivan, and E. F. Hammel (Plenum, New York, 1974), Vol. 3, p.102.
\bibitem{Burlachkov}
L. Burlachkov, Phys Rev. B {\bf 47}, 8056 (1993).
\bibitem{BardSteph}
J. Bardeen and M. J. Stephen, Phys. Rev. {\bf140}, A1197 (1965).
\bibitem{Xu}
Ming Xu, Phys. Rev. B, {\bf 44}, 2713 (1991).
\bibitem{deGennes2}
P. G. deGennes, \textit{Superconductivity of metals and alloys}, W. A. Benjamin Inc., New York, Amsterdam, 1966.
\bibitem{Gough}
S. P. Farrant and C. E. Gough, Phys. Rev. Lett.  {\bf 34}, 943
(1975).
\bibitem{Thouless}
D. J. Thouless, Phys. Rev. Lett. {\bf 34}, 947 (1975)
\bibitem{note}
A possible change of vortex entropy at the peak effect is expected to be
very small, if present, and is certainly not detectable with our technique.
\bibitem{Wheeler}
J. C. Wheeler, Phys. Rev. A {\bf 12}, 267 (1975).
\bibitem{FisherI}
D. S. Fisher, M. P. A. Fisher, and D. A. Huse, Phys. Rev. B {\bf 43}, 130 (1991).
\bibitem{nattII}
T. Nattermann and S. Scheidl, Adv. Phys {\bf 49}, 607 (2000).
\bibitem{Menon}
G. Menon Phys. Rev. B {\bf 65}, 104527 (2002).

\end{thebibliography}

\vspace{0.0cm}
\end{document}